\documentclass[twocolumn,amssymb, nobibnotes, aps, prb, superscriptaddress]{revtex4-1}
\usepackage[pdftex]{graphicx}
\usepackage[ngerman,english]{babel}
\usepackage[ansinew]{inputenc}
\usepackage{natbib}
\usepackage{color}
\usepackage{siunitx}
\usepackage{calc}
\usepackage{lipsum}
\usepackage[colorlinks=true, pdfstartview=FitV, linkcolor=blue,
citecolor=blue, urlcolor=blue]{hyperref}

\definecolor{green2}{rgb}{.0, .58, 0}

\begin{document}
%
\title{Stability of N\'eel-type skyrmion lattice against oblique
  magnetic fields in GaV$_4$S$_8$ and GaV$_4$Se$_8$}
\author{B.~Gross}\affiliation{Department of Physics, University of
  Basel, 4056 Basel, Switzerland}
\author{S.~Philipp}\affiliation{Department of Physics, University of
  Basel, 4056 Basel, Switzerland} \author{K.~Geirhos}
\affiliation{Experimental Physics V, Center for Electronic Correlations and Magnetism, University of Augsburg, 86159 Augsburg, Germany}
\author{A.~Mehlin}\affiliation{Department of Physics, University of
  Basel, 4056 Basel, Switzerland} 
\author{S.~Bord\'acs}\affiliation{Department of Physics, Budapest University of Technology and Economics, 1111 Budapest, Hungary}\affiliation{Hungarian Academy of Sciences, Premium Postdoctor Program, 1051 Budapest, Hungary}
\author{V.~Tsurkan} \affiliation{Experimental Physics V, Center for Electronic Correlations and Magnetism, University of Augsburg, 86159 Augsburg, Germany}
\affiliation{Institute of Applied Physics, MD-2028 Chisinau, Republic of Moldova} 
\author{A. Leonov} \affiliation{Department of Chemistry, Faculty of Science, Hiroshima University Kagamiyama, Higashi Hiroshima, Hiroshima 739-8526, Japan} \author{I.~K\'ezsm\'arki} \affiliation{Experimental Physics V, Center for Electronic Correlations and Magnetism, University of Augsburg, 86159 Augsburg, Germany}
\author{M.~Poggio} \affiliation{Department of Physics, University of Basel, 4056 Basel, Switzerland} \affiliation{Swiss Nanoscience Institute, University of Basel, 4056 Basel, Switzerland}
\begin{abstract} 
  The orientation of N\'eel-type skyrmions in the lacunar spinels
  GaV$_4$S$_8$ and GaV$_4$Se$_8$ is tied to the polar axes of their
  underlying crystal structure through the Dzyaloshinskii-Moriya
  interaction.  In these crystals, the skyrmion lattice phase exists
  for externally applied magnetic fields parallel to these axes and
  withstands oblique magnetic fields up to some critical angle.  Here,
  we map out the stability of the skyrmion lattice phase in both
  crystals as a function of field angle and magnitude using dynamic
  cantilever magnetometry.  The measured phase diagrams reproduce the
  major features predicted by a recent theoretical model, including a
  reentrant cycloidal phase in GaV$_4$Se$_8$.  Nonetheless, we observe
  a greater robustness of the skyrmion phase to oblique fields,
  suggesting possible refinements to the model.  Besides identifying
  transitions between the cycloidal, skyrmion lattice, and
  ferromagnetic states in the bulk, we measure additional anomalies in
  GaV$_4$Se$_8$ and assign them to magnetic states confined to polar
  structural domain walls.
 \end{abstract} 
\maketitle
\section{Introduction}
\label{sec:intro}
The discovery of the nanometer-scale magnetization configurations
known as magnetic skyrmions~\cite{muhlbauer_skyrmion_2009} has spurred
renewed interest in non-centrosymmetric magnets.  The lack of
inversion symmetry in these crystals gives rise to an asymmetric
exchange coupling, known as the Dzyaloshinskii-Moriya interaction
(DMI), which manifests itself in the continuum approximation of the
magnetic order parameter as Lifshitz invariants (LIs) involving first
derivatives of the magnetization $\textbf{M}$ with respect to the
spatial coordinates:
\begin{equation}
\mathcal{L}^{(k)}_{i,j} = M_i \frac{\partial M_j}{\partial x_k} - M_j  \frac{\partial M_i}{\partial x_k}.
\label{Lifshitz}
\end{equation}
Crystal symmetry determines the allowed LIs, i.e.\ a certain
combination of first order derivatives, which -- in competition with
the spin stiffness -- stabilize modulated spin-textures such as
spirals and skyrmions and determine their internal
structure~\cite{bogdanov_thermodynamically_1989,bogdanov_thermodynamically_1994}.
Both skyrmion lattices (SkLs)~\cite{yu_real-space_2010,yu_near_2011}
and isolated skyrmions~\cite{du_edge-mediated_2015} have now been
observed in either bulk or nanostructured noncentrosymmetric crystals.
Their topologically protected spin-texture, which is stable even at
room temperature~\cite{tokunaga_new_2015}, their nanometer-scale size,
and their easy manipulation via electric currents and
fields~\cite{schulz_emergent_2012,jonietz_spin_2010,hsu_electric-field-driven_2017,ruff_polar_2017,fujima_thermodynamically_2017}
make skyrmions a promising platform for information storage and
processing
applications~\cite{sampaio_nucleation_2013,tomasello_strategy_2014}.

Until recently, most investigations in bulk crystals have focused on
Bloch-type skyrmions, in which the local magnetization rotates
perpendicular to the radial direction moving from the skyrmion core to
the far field.  This type of skyrmion has been observed in chiral
cubic helimagnets with B20 structure such as
MnSi~\cite{muhlbauer_skyrmion_2009},
FeGe~\cite{wilhelm_precursor_2011}, or
Cu$_2$OSeO$_3$~\cite{seki_observation_2012}.  Recently, N\'eel-type
skyrmions, in which the local magnetization rotates in a plane
containing the radial direction, have been observed in bulk
GaV$_4$S$_8$,
GaV$_4$Se$_8$~\cite{kezsmarki_neel-type_2015,bordacs_equilibrium_2017,fujima_thermodynamically_2017,white_direct_2018,geirhos_macroscopic_2020,
  butykai_characteristics_2017}, and
GaMo$_4$S$_8$~\cite{butykai_squeezing_2019}.  These materials
crystallize in the cubic lacunar spinel
structure~\cite{ta_phuoc_optical_2013,abd-elmeguid_transition_2004,dorolti_half-metallic_2010,kim_spin-orbital_2014,guiot_avalanche_2013,singh_orbital-ordering-driven_2014,pocha_electronic_2000,ruff_multiferroicity_2015},
which becomes polar below $\sim \SI{45}{\kelvin}$ and the point
symmetry is reduced from T$_d$ to
C$_{3v}$~\cite{kezsmarki_neel-type_2015,
  ruff_multiferroicity_2015,wang_polar_2015,ehlers_skyrmion_2016}. Since
the magnetic order develops in the polar phase, these compounds are
multiferroic.  Furthermore, the skyrmions posses a non-trivial
electric polarization pattern due to the magnetoelectric
effect~\cite{ruff_multiferroicity_2015}, which may enable nearly
dissipation free manipulation of the magnetic order by electric
fields~\cite{wang_polar_2015}.

%
In addition to obvious differences in the spin texture of Bloch- and
N\'eel-type skyrmions, the phase diagrams of cubic helimagnets and
polar skyrmion hosts are markedly different.  In cubic helimagnets,
the LI has an isotropic form
$ w_{\text{DMI}}=\mathbf{M}\cdot \left ( \nabla \times \mathbf{M}
\right )$.  Therefore, the plane of the SkL aligns itself to be nearly
perpendicular to the applied magnetic field, irrespective of the
field's direction.  The isotropic LI also results in a narrow
stability range of for Bloch-type skyrmions in the vicinity of the
magnetic ordering temperature, due to competition with the
longitudinal conical
phase~\cite{muhlbauer_skyrmion_2009,wilhelm_precursor_2011}.  In
contrast, C$_{nv}$ ($n \geq 3$) symmetry only allows an axially
symmetric LI.  Therefore, in polar skyrmion hosts, modulated magnetic
structures with wave vectors perpendicular to the high symmetry, polar
axis are favoured.  In these compounds, the orientation of N\'eel
skyrmions is locked to the polar axis rather than the applied magnetic
field.  Thus, instead of tilting the plane of the SkL, oblique applied
fields distort the configuration of the N\'eel skyrmions and displace
their cores~\cite{leonov_skyrmion_2017}.  This property has two
consequences on the magnetic phase diagram of such materials: 1) the
SkL phase is more robust than in cubic helimagnets, because the
conical phase is suppressed, and 2) its stability range depends on the
direction of the applied field.  In addition to the polar LI, the
second order magnetic anisotropy allowed in this symmetry can also
modify the phase diagram.  In the case of GaV$_4$S$_8$, strong
easy-axis anisotropy~\cite{ehlers_skyrmion_2016} suppresses the
modulated phases at low temperature~\cite{white_direct_2018}, whereas
in GaV$_4$Se$_8$ easy-plane anisotropy helps to stabilize the SkL
phase down to the lowest
temperatures~\cite{bordacs_equilibrium_2017,fujima_thermodynamically_2017,geirhos_macroscopic_2020}.

Here, we use dynamic cantilever magnetometry
(DCM)~\cite{mehlin_stabilized_2015,gross_dynamic_2016,mehlin_observation_2018}
to map the magnetic phase boundaries in GaV$_4$S$_8$ and GaV$_4$Se$_8$
as a function of the strength and orientation of magnetic field.  We
determine the corresponding phase diagrams, which reproduce the major
features predicted by a recent theoretical
model~\cite{leonov_skyrmion_2017}.  The measurements constitute a
direct experimental confirmation of the robustness of N\'eel-type
skyrmions to oblique magnetic fields in two materials with uniaxial
magnetic anisotropy of opposite signs.  In addition to magnetic
transitions between the cycloidal, SkL, and field-polarized
ferromagnetic states, in GaV$_4$Se$_8$, we also observe sharp
anomalies in the torque, which we assign to field-driven
transformations of magnetic states confined to polar domain walls
(DWs).

\section{Dynamic Cantilever Magnetometry}
\label{sec:technique}
In DCM, the sample under investigation is attached to the end of a
cantilever, which is driven into self-oscillation at its resonance
frequency $f$.  Changes in this resonance frequency $\Delta f = f -
f_0$ are measured as a function of the uniform applied magnetic field
$\mathbf{H}$, where $f_0$ is the resonance frequency at $H = 0$.
$\Delta f$ reveals the curvature of the magnetic energy $E_m$ with
respect to rotations about the cantilever oscillation
axis~\cite{mehlin_stabilized_2015,gross_dynamic_2016}:
\begin{equation}
  \Delta f = \frac{f_0}{2 k_0 l_e^2}\left( \left.\frac{\partial^2E_m}{\partial\theta_c^2}\right|_{\theta_c=0}\right),
  \label{eq:deltaf}
\end{equation}
where $k_0$ is the cantilever's spring constant, $l_e$ its effective
length, and $\theta_c$ its angle of oscillation.  Measurements of this
magnetic curvature are particularly useful for identifying magnetic
phase transitions~\cite{mehlin_stabilized_2015}, since -- just as the
magnetic susceptibility -- it should be discontinuous for both first
and second order phase transitions~\cite{modic_resonant_2018}.

\begin{figure*}[t]
\includegraphics{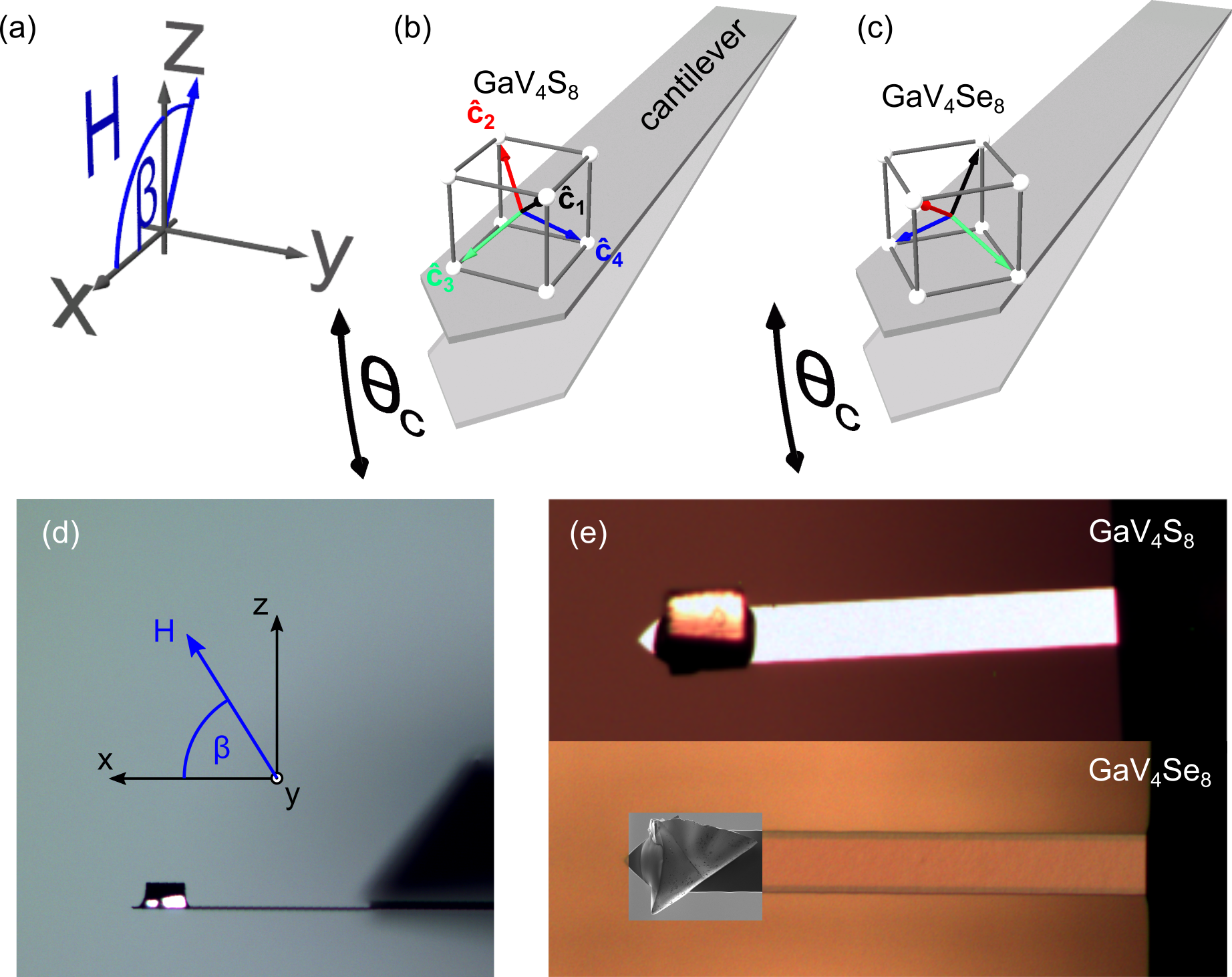}
\caption{Schematics of the measurement setup. (a) shows the coordinate
  system and the definition of $\beta$ as the angle between
  $\mathbf{H}$ and $\mathbf{\hat{x}}$. (b) and (c) show the
  cantilever, its oscillation angle $\theta_c$, and the crystalline
  axes of the measured sample.  Black, red, green, and blue lines
  correspond to the four $\mathbf{\hat{c}_i}$.  (d) shows the
  orientation of $\mathbf{H}$ with respect to an optical image of a
  sample and cantilever.  (e) Composite optical and scanning electron
  micrographs of the measured GaV$_4$S$_8$ and GaV$_4$Se$_8$ samples
  mounted on their respective cantilevers.}
\label{fig:setup}
\end{figure*}

DCM measurements are carried out in a vibration-isolated closed-cycle
cryostat.  The pressure in the sample chamber is less than
$10^{-6}$~mbar and the temperature can be stabilized between 4 and
\SI{300}{\kelvin}.  Using an external rotatable superconducting
magnet, magnetic fields up to \SI{4.5}{\tesla} can be applied along
any direction spanning \SI{120}{\degree} in the plane of cantilever
oscillation, as shown in Fig.~\ref{fig:setup}.  $\mathbf{\hat{x}}$ in
our coordinate system is defined by the cantilever's long axis, while
$\mathbf{\hat{y}}$ coincides with its axis of oscillation.  $\beta$ is
the angle between $\mathbf{H}$ and $\mathbf{\hat{x}}$ in the
$xz$-plane.  The cantilever's motion is read out using a optical fiber
interferometer using \SI{100}{\nano\watt} of laser light at
\SI{1550}{\nano\meter}~\cite{rugar_improved_1989}.  A piezoelectric
actuator mechanically drives the cantilever at $f$ with a constant
oscillation amplitude of a few tens of nanometers (corresponding to
oscillation angles of tens of microradians) using a feedback loop
implemented by a field-programmable gate array.  This process enables
the fast and accurate extraction of $f$ from the cantilever deflection
signal as well as providing a measure of the dissipation $\Gamma$,
which described the system's rate of energy loss:
$d E / d t = - \Gamma l_e^2 \dot{\theta_c}^2$.  In order to maintain a
constant oscillation amplitude, the cantilever must be driven with a
force $F = \Gamma l_e \dot{\theta_c}$, such that any losses due to
dissipation are compensated.  The voltage amplitude used to drive the
piezoelectric actuator is therefore proportional to
$\Gamma = \Gamma_0 + \Gamma_m$ where $\Gamma_0$ is the cantilever's
intrinsic mechanical dissipation at $H = 0$ and $\Gamma_m$ represents
magnetic losses.  Given that $\Gamma_m$ reflects the sample's magnetic
relaxation, $\Gamma$ should undergo abrupt changes at magnetic phase
transitions.  We therefore use both measurements of the magnetic
curvature and dissipation, combined with knowledge from other
measurements~\cite{kezsmarki_neel-type_2015,bordacs_equilibrium_2017,fujima_thermodynamically_2017,white_direct_2018,geirhos_macroscopic_2020},
to map the low-temperature magnetic phase diagrams of GaV$_4$S$_8$ and
GaV$_4$Se$_8$ as a function of $\mathbf{H}$.
\section{Samples}
\label{sec:samples}
Single crystals of GaV$_4$S$_8$ and GaV$_4$Se$_8$ are grown by a
chemical transport reaction method using iodine as a transport
agent~\cite{kezsmarki_neel-type_2015}.  X-ray diffraction measurements
of both sample materials show impurity-free
single-crystals~\cite{ruff_multiferroicity_2015}.  For the DCM
measurement, we attach individual crystals of GaV$_4$S$_8$ and
GaV$_4$Se$_8$, which are a few tens of micrometers in size, to the
ends of commercial Si cantilevers (Nanosensors\texttrademark TL-cont)
using non-magnetic epoxy, as shown in Fig.~\ref{fig:setup}.  These
cantilevers are \SI{440}{\micro\meter}-long,
\SI{50}{\micro\meter}-wide, and \SI{2.3}{\micro\meter}-thick.
Unloaded, they have resonance frequencies of about
\SI{16}{\kilo\hertz}, quality factors around $5 \times 10^5$, and
spring constants of \SI{300}{\milli\newton/\meter}.  Due to the
additional mass of the samples, the resonance frequency of a loaded
cantilever shifts to around \SI{3}{\kilo\hertz}.

Both samples are attached near the free end of the cantilever with the
$(001)$ surface pressed flat against the Si surface.
The orientation of the GaV$_4$S$_8$ and GaV$_4$Se$_8$ samples differs
and can be roughly estimated from optical and scanning electron
microscope (SEM) images.  The resultant direction of each sample's
crystalline axes with respect to the cantilever is shown in
Fig.~\ref{fig:setup}: specifically the approximate orientation of the
four cubic $\langle 111 \rangle$ axes $\mathbf{\hat{c}_i}$
($i = 1, 2, 3, 4$) is shown in black, red, green, and blue.

Both GaV$_4$S$_8$ and GaV$_4$Se$_8$ undergo a Jahn-Teller structural
phase transition from a non-centrosymmetric cubic to a rhombohedral
structure at $\SI{44}{\kelvin}$ and $\SI{42}{\kelvin}$,
respectively~\cite{pocha_electronic_2000,ruff_multiferroicity_2015,fujima_thermodynamically_2017,ruff_polar_2017}.
The transition is characterized by a stretching of the cubic unit cell
along one of the four cubic body diagonals $\mathbf{\hat{c}_i}$,
resulting in four different structural domains.  The rhombohedral
distortion also gives rise to polarization along $\mathbf{\hat{c}_i}$,
making these the polar axes of the system.  The multi-domain state is
composed of sub-micrometer-thick sheets of these four different
rhombohedral polar domains, which we label
P$_i$~\cite{butykai_characteristics_2017,geirhos_macroscopic_2020}.
The polar axis $\mathbf{\hat{c}_i}$ also corresponds to the axis of
magnetic anisotropy in the respective rhombohedral domain state.
In GaV$_4$S$_8$, the uniaxial anisotropy is of easy-axis type, while
in GaV$_4$Se$_8$ it is of easy-plane
type~\cite{kezsmarki_neel-type_2015,bordacs_equilibrium_2017,ehlers_skyrmion_2016}.
In both materials, measurements indicate the presence of modulated
magnetic phases including a cycloidal (Cyc) state, a N\'eel-type SkL,
and a field polarized ferromagnetic (FM)
phase~\cite{kezsmarki_neel-type_2015,bordacs_equilibrium_2017}.  The
population of multiple rhombohedral domains at low temperature
complicates the determination of the magnetic phase diagram, because
for any given orientation of the applied field $\mathbf{H}$, there can
be up to four different angles, $\alpha_i$, between $\mathbf{H}$ and
$\mathbf{\hat{c}_i}$.  As a result, for an arbitrary orientation of
$\mathbf{H}$, a single phase transition can appear at up to four
different values of $H$, depending on the projections of $\mathbf{H}$
on each $\mathbf{\hat{c}_i}$.  Although the application of a large
electric field upon cooling through the structural phase transition
has been shown to polarize GaV$_4$S$_8$ and GaV$_4$Se$_8$ samples such
that only a single domain is
populated~\cite{fujima_thermodynamically_2017, ruff_polar_2017}, it is
practically challenging to apply such fields in a DCM apparatus.
\section{Measurements}
\label{sec:measurements}
\subsection{GaV$_4$S$_8$}
\label{subsec:GaVS}
\begin{figure*}[t]
  \includegraphics[width=16cm]{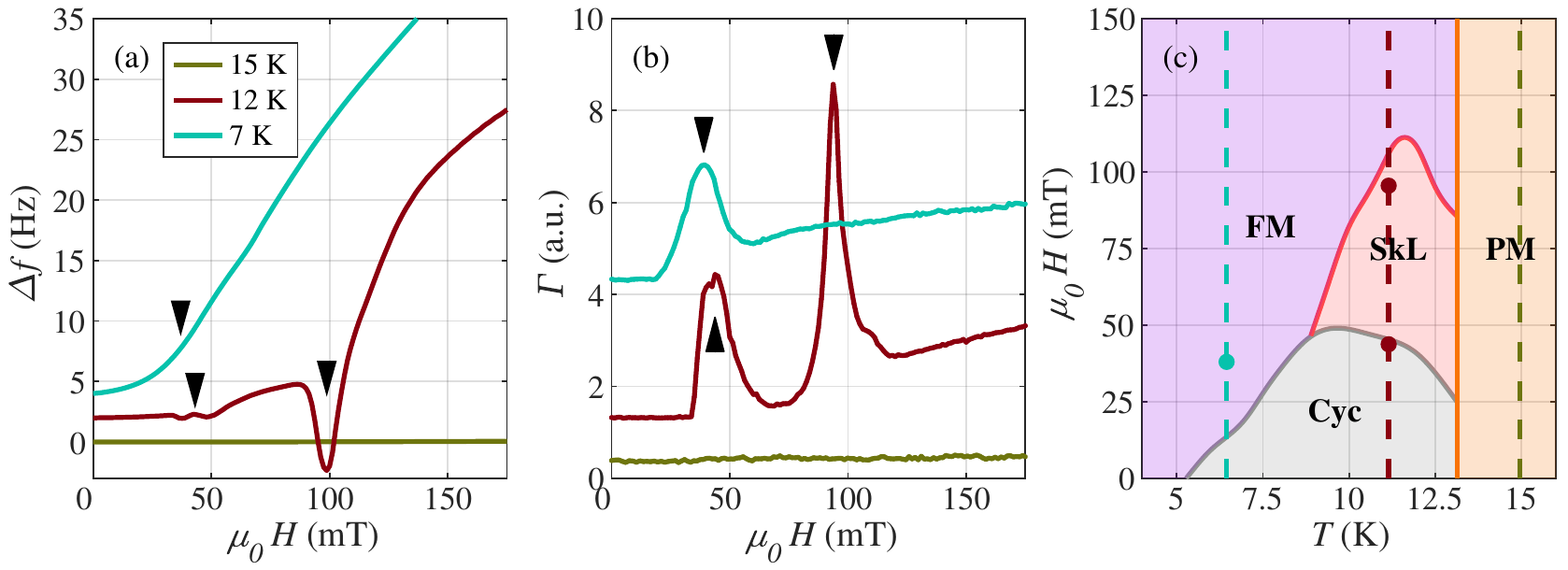}
  \caption{Temperature and field dependence of magnetic phase
    transitions measured by DCM in GaV$_4$S$_8$.  (a) DCM measurements
    of (a) $\Delta f(H)$ and (b) $\Gamma(H)$ taken at $T = 7$, $12$,
    and $\SI{15}{\kelvin}$ in cyan, maroon, and brown,
    respectively. Curves are shifted for better visibility.
    $\beta = 0$, i.e.\ approximately $\mathbf{H} \parallel [100]$.
    Arrows indicate features corresponding to phase transitions.  (c)
    Sketch of the expected magnetic phase diagram as a function of
    temperature and applied field for
    $\mathbf{H} \parallel [100]$~\cite{kezsmarki_neel-type_2015}.
    Color-coded dashed lines and points correspond to temperatures and
    measured features in (a) and (b).}
\label{fig:exampleGaVS}
\end{figure*}
Fig.~\ref{fig:exampleGaVS} shows DCM measurements of $\Delta f(H)$ and
$\Gamma_m(H)$ in GaV$_4$S$_8$ for different temperatures $T$.  Data shown in
Fig.~\ref{fig:exampleGaVS}~(a) and (b) are collected with $\mathbf{H}$
aligned along the cantilever's long axis ($\beta = 0$), i.e.\
approximately $\mathbf{H} \parallel [100]$.  In this configuration,
the angles $\alpha_i$ between $\mathbf{H}$ and the four
$\mathbf{\hat{c}_i}$ are the same within the precision of the sample
orientation, i.e.\ within a few degrees.  Consequently, each magnetic
phase transition should occur at a similar value of $H$ for each
domain.  In this particularly simple case, we compare $\Delta f(H)$
and $\Gamma(H)$ at different temperatures to the corresponding
magnetic phase diagram measured by K\'ezsm\'arki et
al.~\cite{kezsmarki_neel-type_2015} and shown schematically in
Fig.~\ref{fig:exampleGaVS}~(c).  Where metamagnetic transitions are
expected, they manifest themselves as dips in $\Delta f(H)$ and peaks
in $\Gamma_m(H)$.  At $T = \SI{12}{\kelvin}$, the two features at 45
and \SI{100}{\milli\tesla} (indicated by arrows) correspond to the
Cyc-to-SkL and the SkL-to-FM phase transitions, respectively.  The
double dip (peak) feature in $\Delta f (H)$ ($\Gamma(H)$) comes from
the imperfect alignment of the sample's crystalline axes with the
coordinate system of our measurement setup, resulting in a difference
in $\alpha_i$ for each domain.  At $T = \SI{7}{\kelvin}$ only one
feature is found, corresponding to the Cyc-to-FM transition, while at
$T = \SI{15}{\kelvin}$, which is above the magnetic ordering
temperature, no features are observed.
\begin{figure*}[t]
  \includegraphics[width=16cm]{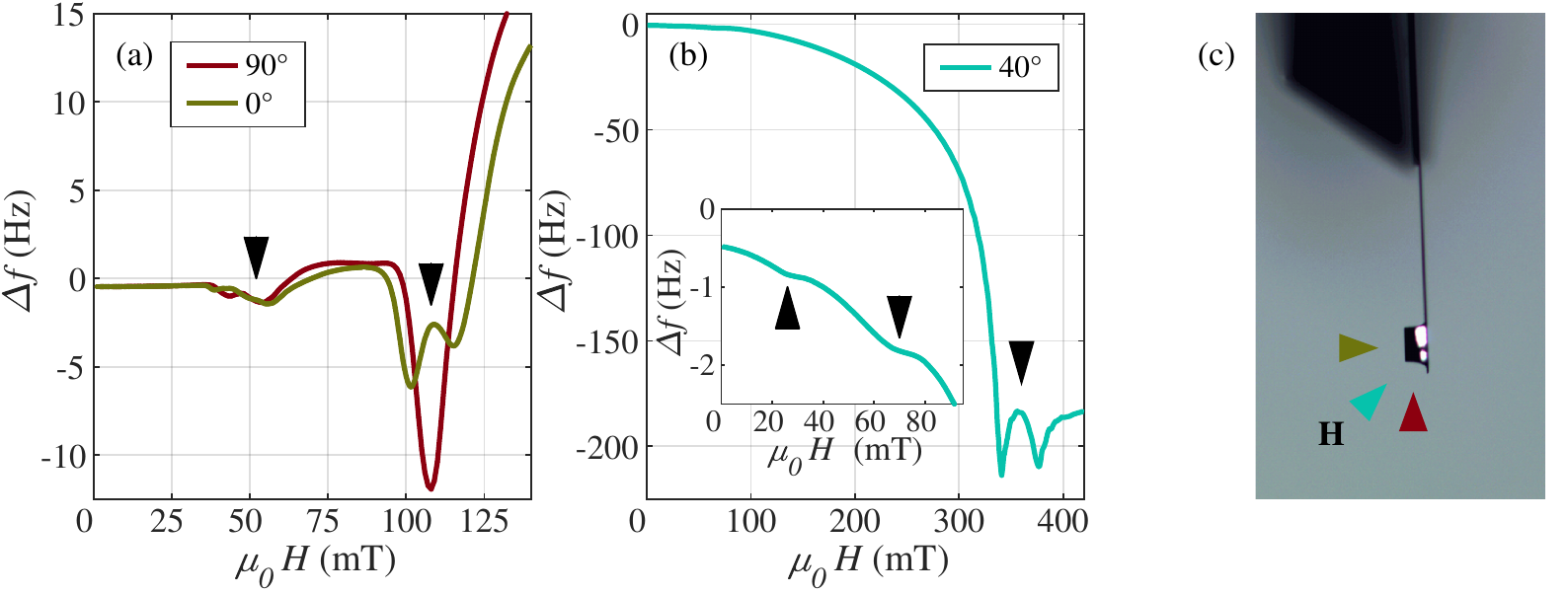}
  \caption{Angular dependence of magnetic phase transitions measured
    by DCM in GaV$_4$S$_8$.  (a), (b) $\Delta f(H)$ at $T =
    \SI{11}{\kelvin}$ for at $\beta = 0$, $40$, and $\SI{90}{\degree}$
    in maroon, cyan, and brown, respectively.  Arrows indicate
    features corresponding to phase transitions.  Inset: zoomed view
    of the low-field region.  (c) Schematic diagram showing the three
    measured orientations relative to the sample-loaded cantilever.}
\label{fig:exampleGaVS2}
\end{figure*}

$\mathbf{H}$ is rotated approximately in the (010) plane such that, in
general, by changing $\beta$, we change each $\alpha_i$ differently.
As a result, the number of features related to phase transitions and
the fields at which they occur can also change.  The dependence that
we observe is consistent with the orientation of our sample and
previous measurements by K\'ezsm\'arki et al.  In particular, we note
that because of the crystal's alignment and its cubic symmetry, the
measured curves should repeat themselves upon rotating $\beta$ by
\SI{90}{\degree}.  This periodic behavior can be seen in
Fig.~\ref{fig:exampleGaVS2}~(a), where two DCM curves with $\beta = 0$
and \SI{90}{\degree} nearly overlap; differences, including the
splitting of the dips in $\Delta f (H)$ into two dips, are again
related to the slight misalignment of the sample's crystalline axes
with respect to the applied field, resulting in slightly different
$\alpha_i$ for each domain.  In the curve taken with
$\beta = \SI{40}{\degree}$ (approximately
$\mathbf{H} \parallel [101]$) shown in Fig.~\ref{fig:exampleGaVS2}
(b), we observe four features.  The features observed at 35 and
\SI{60}{\milli\tesla} are the Cyc-to-SkL and the SkL-to-FM phase
transitions, respectively, also observed by K\'ezsm\'arki et al.
These transitions correspond to the P$_4$ and P$_1$ domains (blue and
black in Fig.~\ref{fig:setup}) with $\alpha_4 = \SI{31.7}{\degree}$ and
$\alpha_1 = \SI{39.2}{\degree}$.  The two transitions at
\SI{320}{\milli\tesla} and \SI{370}{\milli\tesla} correspond to the
Cyc-to-FM transitions in the P$_3$ and P$_2$ domains (green and red
in Fig.~\ref{fig:setup}), where $\alpha_3 = \SI{84.5}{\degree}$ and
$\alpha_2 = \SI{88.8}{\degree}$.  As before, the mismatches
$\alpha_4 \neq \alpha_1$ and $\alpha_3 \neq \alpha_2$ and the
resulting pair of split features are due to the crystal's imperfect
alignment with the applied field.

\begin{figure}[t]
\includegraphics[]{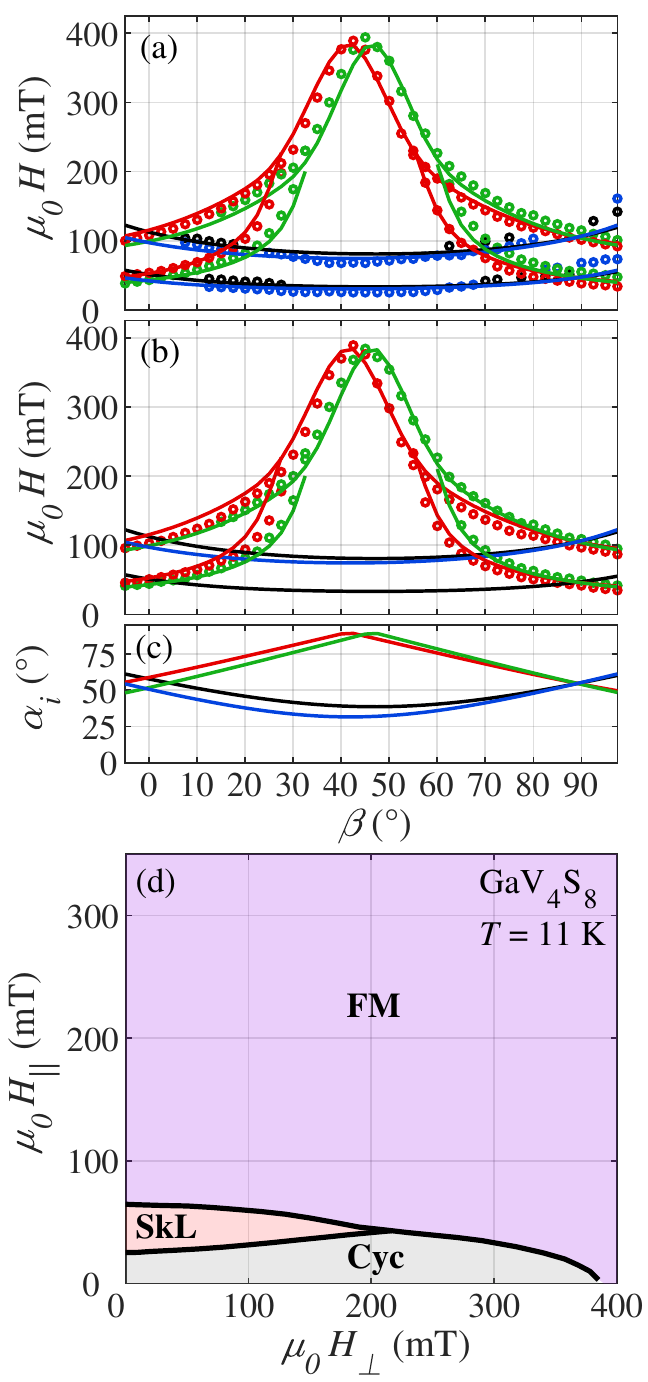}
\caption{Magnetic phase transitions measured in GaV$_4$S$_8$ at
  $T = \SI{11}{\kelvin}$.  Features extracted from DCM measurement of
  (a) $\Delta f (H)$ and (b) $\Gamma (H)$ are plotted as open circles
  as a function of $\beta$.  Black, red, green, and blue circles
  correspond to domains P$_1$, P$_2$, P$_3$, and P$_4$, respectively.
  Color-coded lines indicate phase boundaries for each domain
  according to the phase diagram in (d). (c) Angle $\alpha_i$ between
  the corresponding $\mathbf{\hat{c}_i}$ and the external field
  $\mathbf{H}$ vs.\ $\beta$ for all four rhombohedral domains, using
  the same color code as in (a) and (b). (d) Best-fit magnetic phase
  diagram for single-domain GaV$_4$S$_8$ as a function of field
  applied perpendicular and parallel to the axis of symmetry.}
\label{fig:GaVS}
\end{figure}

Using the measured features in $\Delta f(H)$ and $\Gamma_m(H)$, we map
the magnetic phase transitions of GaV$_4$S$_8$ as a function of $H$
and $\beta$.  After initializing the system with a large external
field $H = \SI{1}{\tesla}$, DCM measurements are made by stepping $H$
toward zero at a fixed $\beta$ and $T$.  The angular dependence over
the range $\SI{-5}{\degree} < \beta < \SI{100}{\degree}$ is recorded
at $T = \SI{11}{\kelvin}$ by changing $\beta$ in steps of
\SI{2.5}{\degree} and repeating the measurement.  We plot the features
identified in these measurements as open circles in
Figs~\ref{fig:GaVS}~(a) and (b).  By comparing our data taken for a
few magnetic field orientations with the phase diagram reported by
K\'ezsm\'arki et al.~\cite{kezsmarki_neel-type_2015}, we assign each
feature to a certain type of transition (i.e.\ Cyc-to-FM, Cyc-to-SkL,
SkL-to-FM) occurring in a certain domain state (P$_1$, P$_2$, P$_3$,
P$_4$).

Next, we determine the dependence of the phase boundaries on the
orientation of the magnetic field with respect to the axis of the
uniaxial magnetic anisotropy.  The measured signatures shown as open
circles in Figs.~\ref{fig:GaVS}~(a) and (b) can be fit by assuming
that each of the four rhombohedral domains of GaV$_4$S$_8$ obeys the
magnetic phase diagram shown in (d), plotted as a function of
$H_\parallel$ and $H_\perp$, the components of $\mathbf{H}$ parallel
and perpendicular to the rhombohedral axis $\mathbf{\hat{c}_i}$,
respectively.  A feature in $\Delta f$ and $\Gamma$ observed at
certain $H$ and $\beta$ corresponds to a transition of a particular
domain P$_i$ for a field of magnitude $H$ and angle $\alpha_i$ with
respect to $\mathbf{\hat{c}_i}$, as shown in Fig.~\ref{fig:GaVS}~(c).
The magnitude $H$ and the angle $\alpha_i$ at which each feature
occurs, correspond to a point on a phase boundary in the diagram of
Fig.~\ref{fig:GaVS}~(d), through $H_\parallel = H \cos{\alpha_i}$ and
$H_\perp = H \sin{\alpha_i}$.  This phase diagram reflects the general
form suggested by Leonov and
K\'ezsm\'arki~\cite{leonov_skyrmion_2017}. Phase boundaries
corresponding to the diagram are also plotted as a function of $\beta$
and $H$ in Figs.~\ref{fig:GaVS}~(a) and (b) to show their agreement
with the measurements.  They appear as solid lines, which are
color-coded according to the domain to which they belong.  An Euler
rotation of the crystal (-5.0, 0.2 and \SI{10.0}{\degree}) with
respect to ideal configuration, shown in Fig.~\ref{fig:setup}~(b), is
required such that the phase boundaries corresponding to the different
domain states collapse onto the single
boundary diagram of Fig.~\ref{fig:GaVS}~(d).

The agreement between the measured features and fit phase boundaries
allows us to eliminate complications arising from the multi-domain
nature of the crystal and, thus, to extract a the general magnetic
phase diagram of GaV$_4$S$_8$ as function of field applied parallel
and perpendicular to the anisotropy axis.  The position of the
intersection between the different phase transitions in
Fig.~\ref{fig:GaVS} (d) shows that the SkL phase in GaV$_4$S$_8$
persists in oblique fields up to a threshold angle as large as
$\alpha_{\text{max}} = \SI{77}{\degree}$.  For larger $\alpha$, the
cycloidal state directly transforms to the ferromagnetic state upon
increasing $H$.  The extent of the SkL phase shows stronger stability
against fields applied perpendicular to the anisotropy axis (up to
$H_\perp = \SI{200}{\milli\tesla}$) than fields applied parallel (up
to $H_\parallel = \SI{65}{\milli\tesla}$).  This critical angle
$\alpha_{\text{max}}$ is larger than predicted by Leonov and
K\'ezsm\'arki~\cite{leonov_skyrmion_2017}.
%
%
\subsection{GaV$_4$Se$_8$}
\label{subsec:GaVSe}
\begin{figure*}
\includegraphics[]{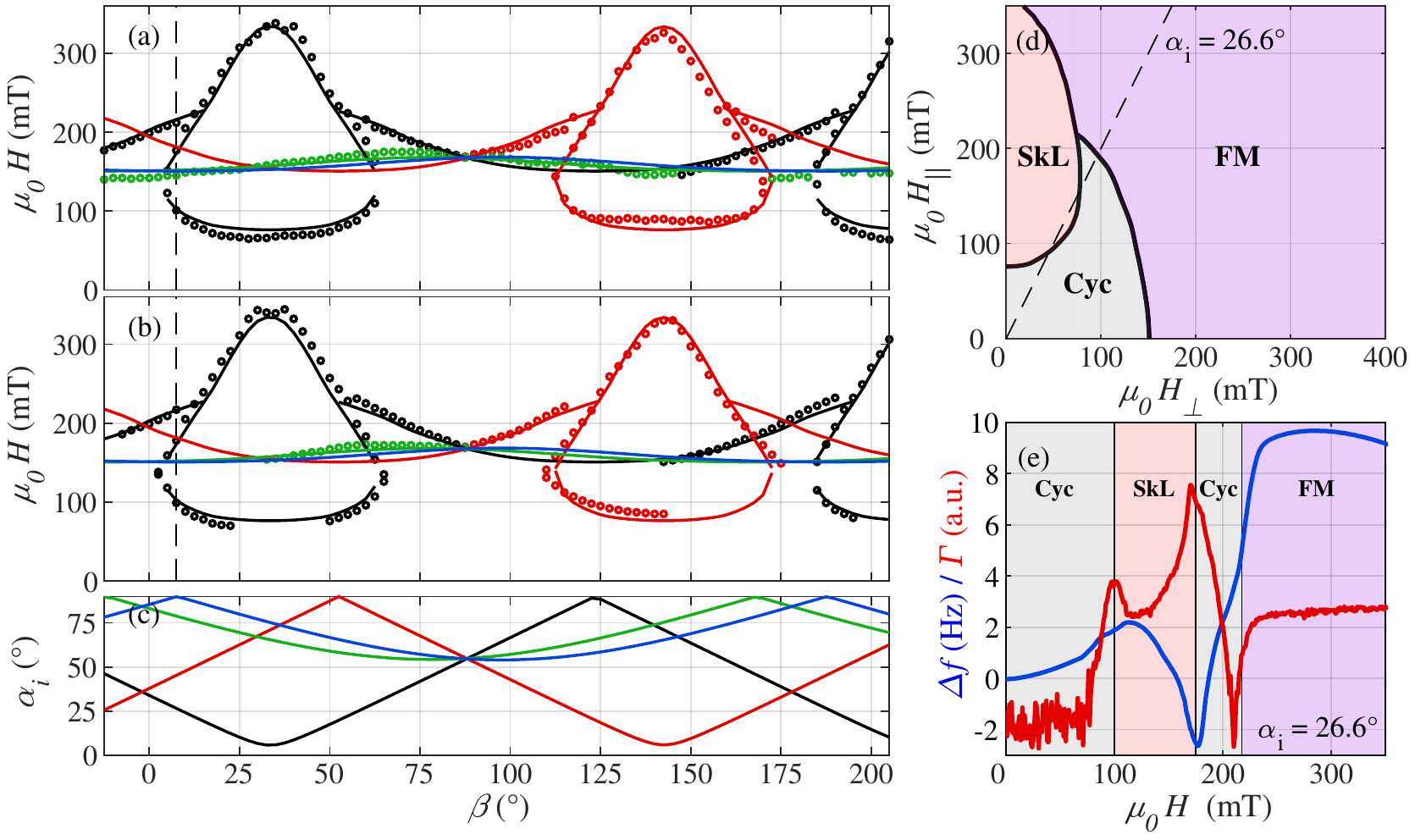}
\caption{Magnetic phase transitions measured in GaV$_4$Se$_8$ at
  $T = \SI{12}{\kelvin}$.  Transitions extracted from DCM measurement
  of (a) $\Delta f (H)$ and (b) $\Gamma (H)$ are plotted as open
  circles as a function of $\beta$.  Black, red, green, and blue
  circles correspond to transitions for domain P$_1$, P$_2$, P$_3$,
  and P$_4$, respectively.  Color-coded lines correspond to phase
  boundaries for the each color-coded domain as indicated by lines in
  the phase diagram (d). (c) Angle $\alpha_i$ between corresponding
  polar axis and the external field $\mathbf{H}$ vs $\beta$ for all
  four rhombohedral domains, using the same color code as in (a) and
  (b). (d) Best-fit magnetic phase diagram for single-domain
  GaV$_4$Se$_8$ as a function of field applied perpendicular and
  parallel to the axis of symmetry.  (e) DCM measurement of
  $\Delta f(H)$ for $\alpha_i = \SI{26.6}{\degree}$
  ($\beta = \SI{7.5}{\degree}$) showing the reentrant Cyc phase. These
  measurements corresponds to line-cuts along the dashed vertical
  lines in (a) and (b) and the dashed diagonal line in (d).}
\label{fig:GaVSe}
\end{figure*}

We apply the same experimental procedure to explore the magnetic phase
diagram of GaV$_4$Se$_8$.  In this case, $\mathbf{H}$ is rotated
approximately in the $(1\bar{1}0)$ plane.  Figs.~\ref{fig:GaVSe}~(a)
and (b) show the angular dependence of the features, as extracted from
measurements of $\Delta f(H)$ and $\Gamma_m(H)$ at
$T = \SI{12}{\kelvin}$.  Using previous measurements made by Bord\'acs
et al.\ along particular crystalline
directions~\cite{bordacs_equilibrium_2017}, as well as neutron
diffraction data by Geirhos et al.\ \cite{geirhos_macroscopic_2020}
for guidance, we assign each feature to a transition between Cyc, SkL,
or FM states for a certain domain and color-code it accordingly.

Once again, the measured features are shown as open circles in
Figs.~\ref{fig:GaVSe}~(a) and (b) and can be fit by assuming that each
of the four rhombohedral domains obeys a single magnetic phase diagram
shown in (d).  The magnitude of the applied field $H$ and its angle
$\alpha_i$ with respect to the assigned domain's rhombohedral axis
$\mathbf{\hat{c}_i}$ put each feature on one of the phase boundaries
depicted in Fig.~\ref{fig:GaVSe}~(d).  Phase boundaries corresponding
to the phase diagram are plotted in Figs.~\ref{fig:GaVSe}~(a) and (b)
for comparison with the measured data.  They appear as solid lines,
which are color-coded according to the domain.  Similarly to
GaV$_4$S$_8$, the overall form of the phase diagram agrees with that
suggested by Leonov and K\'ezsm\'arki~\cite{leonov_skyrmion_2017},
although there are minor quantitative differences between our results
and the theoretical predictions. Note that the rotation plane of
$\mathbf{H}$, approximately $(1\bar{1}0)$, contains
$\mathbf{\hat{c}}_1$ and $\mathbf{\hat{c}}_2$, but not
$\mathbf{\hat{c}}_3$ and $\mathbf{\hat{c}}_4$.  An Euler rotation of
the crystal (-14, -1 and \SI{7}{\degree}) with respect to ideal
configuration, shown in Fig.~\ref{fig:setup}~(c), is required such
that the phase boundaries corresponding to the different domain states
(P$_1$, P$_2$, P$_3$, P$_4$) collapse onto the single boundary diagram
of Fig.~\ref{fig:GaVSe}~(d).
%
We find additional anomalies in both $\Delta f(H)$ and $\Gamma_m(H)$,
that cannot be ascribed to the boundaries between the Cyc, SkL, and FM
phases.  We suspect that these anomalies originate from the formation
of magnetic textures localized at structural DWs, as discussed in
section~\ref{subsec:DW}.

For the black and red domains, which are the only two experiencing
sufficient $H_\parallel$ to reach the SkL phase, the boundaries of the
SkL state appear as prominent rain-drop-like shapes in
Figs~\ref{fig:GaVSe}~(a) and (b).  From the intersection of the SkL
with the Cyc phase boundary in (d), we extract a threshold angle
$\alpha_{\text{max}} = \SI{31}{\degree}$ for the SkL phase in
GaV$_4$Se$_8$ at $T = \SI{12}{\kelvin}$.  Contrary to GaV$_4$S$_8$,
the extent of the SkL phase shows stronger stability against fields
applied parallel to the anisotropy axis (up to
$H_\parallel = \SI{340}{\milli\tesla}$) than fields applied
perpendicular (up to $H_\perp = \SI{75}{\milli\tesla}$).  Furthermore,
we note the presence of a reentrant Cyc phase for angles
\SI{19}{\degree} $ < \alpha_i < $ \SI{30}{\degree}, as predicted by
Leonov and K\'ezsm\'arki~\cite{leonov_skyrmion_2017}.  For this range
of $\alpha_i$, two successive first-order phase transitions from Cyc
to SkL and back occur as a function of increasing field. The signature
of this behavior in DCM is shown in Fig.~\ref{fig:GaVSe}~(e).

\subsection{Magnetic States Confined to Domain Walls in GaV$_4$Se$_8$}
\label{subsec:DW}
Geirhos et al.\ observed anomalies in various macroscopic
thermodynamic properties of GaV$_4$Se$_8$, emerging exclusively in
crystals with polar multi-domain structure.  They suggest a possible
scenario for the formation of magnetic states at the structural DWs of
the lacunar spinel GaV$_4$Se$_8$~\cite{geirhos_macroscopic_2020}.
Magnetic interactions change stepwise at the DWs and spin textures
with different spiral planes, hosted by neighboring domains, need to
be matched there. This can, for example, lead to conical magnetic
states at the DWs with a different closing field magnitude than bulk
magnetic states.  Here, we adopt and modify this model in order to
analyze its applicability to anomalies observed in our DCM
measurements of GaV$_4$Se$_8$, which cannot be assigned to bulk
magnetic phase transitions.

\begin{figure}
\includegraphics[width=8.5cm]{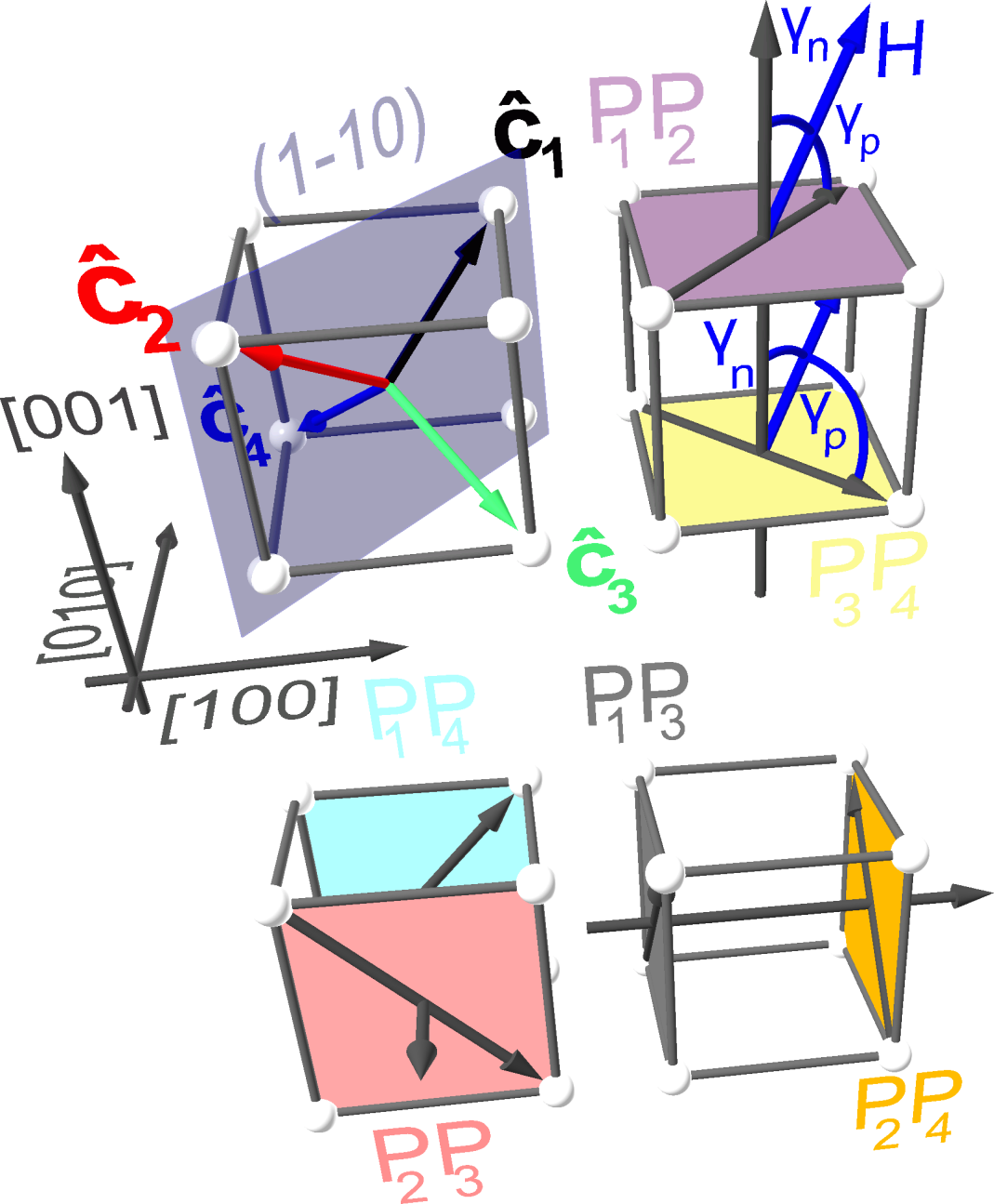}
\caption{Schematic for understanding the orientation of the 6
  different domain walls types. Top left: Directions of the four
  possible polar axes, P$_1$-P$_4$, which are the axes of magnetic
  anisotropy within the corresponding domains. The transparent blue
  plane indicates the approximate plane of rotation of the external
  magnetic field.  Top right: Mechanically compatible and charge
  neutral DWs separating P$_1$ and P$_2$ domains are parallel to the
  $(001)$ plane, just as DWs between P$_3$ and P$_4$ domains. The
  former and latter DWs are referred to as P$_1$P$_2$ and P$_3$P$_4$,
  respectively. $\gamma_n$ ($\gamma_p$), the angle between
  $\mathbf{H}$ and $\mathbf{\hat{c}}_i+\mathbf{\hat{c}}_j$
  ($\mathbf{\hat{c}}_i-\mathbf{\hat{c}}_j$), is shown for both DW
  pairs.  Bottom: The other two pairs of DWs sharing the same
  orientation. The normal vector of the corresponding planes and their
  labels are indicated for the three cases, as well as the difference
  vector $\mathbf{\hat{c}}_i-\mathbf{\hat{c}}_j$, unique to each DW
  type.}
\label{fig:DWtypes}
\end{figure}

In the rhombohedral phase of the studied lacunar spinels, mechanically
compatible and charge neutral DWs are normal to
$\mathbf{\hat{c}}_i+\mathbf{\hat{c}}_j$, the sum of the two polar
directions of the domain states P$_i$ and P$_j$ separated by the DW,
as shown in Fig.~\ref{fig:DWtypes}~\cite{geirhos_macroscopic_2020,
  neuber_architecture_2018, butykai_characteristics_2017}. For
example, mechanically and electrically compatible DWs connecting a
P$_1$ (black) and a P$_2$ (red) domain are parallel to (001) planes,
cf. Fig.~\ref{fig:DWtypes}.
The same is true for DWs between P$_3$ (green) and P$_4$ (blue)
domains. 

For an arbitrary orientation of the external magnetic field, magnetic
states confined to DWs with different orientations are expected to
undergo field-induced transitions, similarly to the bulk (in-domain)
magnetic states.  However, in this case the situation is more complex:
The stability of the magnetic states confined to DWs is determined by
the orientation of the field with respect to the magnetic anisotropy
axes of adjacent domains and to the DW itself.

\begin{figure*}
\includegraphics[width=12cm]{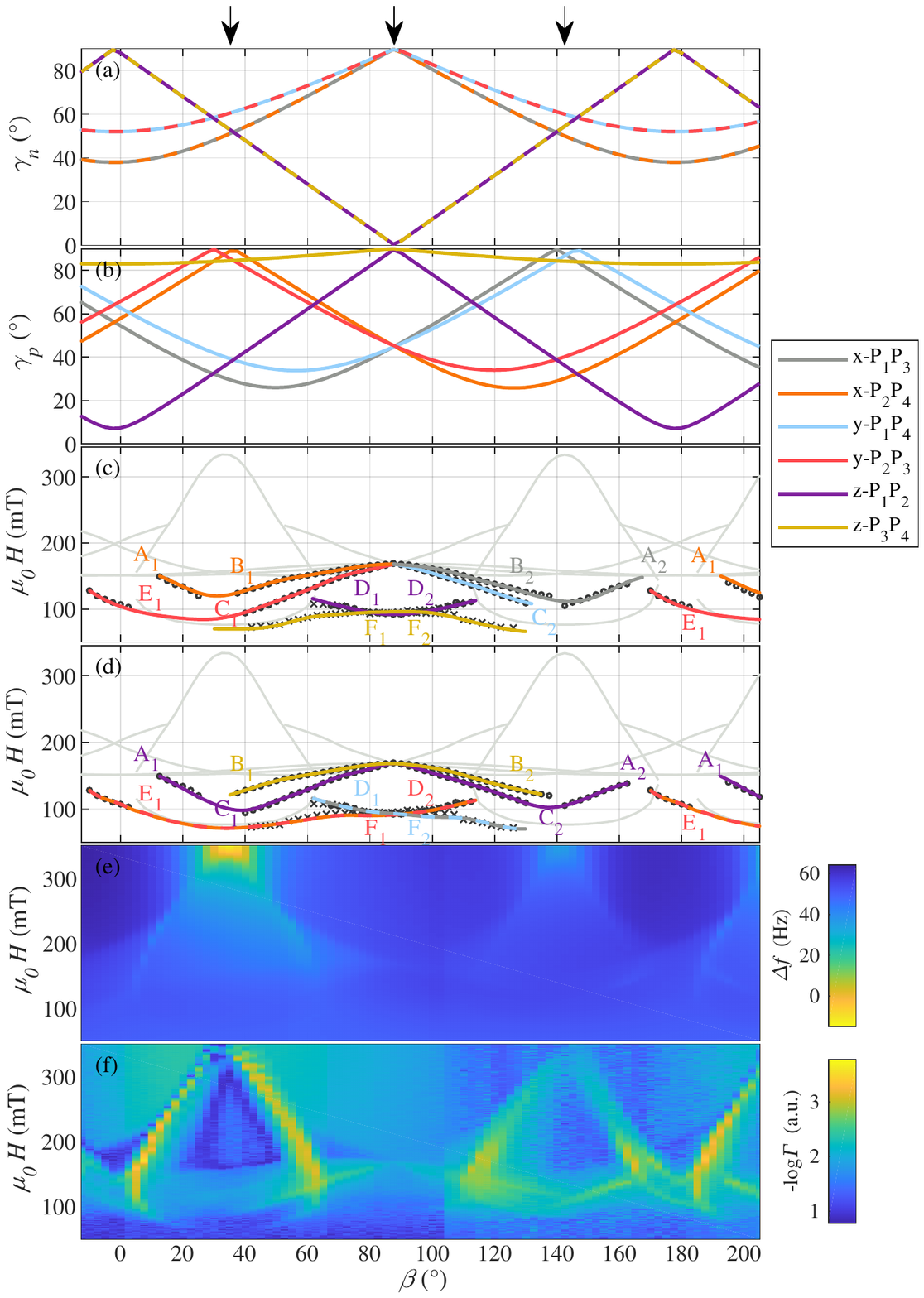}
\caption{Anomalies in $\Delta f(H)$ and $\Gamma(H)$ assigned to transitions of DW rather than bulk magnetic states.  Arrows at the top indicate from left to right the approximate angle $\beta$ corresponding to the $[111]$, $[001]$ and $[11-1]$ directions, respectively.  (a) Angle $\gamma_n$ between the normal vector of a DW and $\mathbf{H}$ plotted against $\beta$. The color of the dashed lines shows their correspondence to a DW type in the legend. (b) Angle $\gamma_p$ between the vector formed by the sum of the polar axis vectors of the two adjacent domains of a DW and $\mathbf{H}$ plotted against $\beta$. (c) Transitions extracted from both $\Delta f$ and $\Gamma$ that are not assigned to a domain transition (dark gray circles). Crosses show transitions extracted from magnetoelectric measurements~\cite{geirhos_macroscopic_2020}, scaled by about 0.9 to match the DCM data.  Colored lines show the suggested assignment of the transitions to DW types as denoted in the legend. (d) Same data as in (c) with a different assignment of transitions.  Color map of (e) $\Delta f (H, \beta)$ and (f) $-\log_{10} \Gamma_m (H, \beta)$.}
\label{fig:DWdata}
\end{figure*}

It is reasonable to assume, that the angle, $\gamma_n$, between
$\mathbf{H}$ and the normal of the DW planes, given by
$\mathbf{\hat{c}}_i+\mathbf{\hat{c}}_j$, plays a decisive role in
setting the angular range, across which confined states are stable.
This leads to three pairs of DWs, as shown in Fig.~\ref{fig:DWtypes},
each sharing the same $\gamma_n$ for a given $\mathbf{H}$.  For DWs in
a pair, however, the relative orientation between the magnetic
anisotropy axes of the two domains involved and $\mathbf{H}$ is not
the same.  For example, consider the P$_1$P$_2$/P$_3$P$_4$ pair: the
rotation plane of $\mathbf{H}$ ($1\bar{1}0$) contains the anisotropy
axes of P$_1$ and P$_2$, but not the anisotropy axes of P$_3$ and
P$_4$; they span $\SI{54}{\degree}$ with this plane.  We therefore
introduce another angle, $\gamma_p$, between $\mathbf{H}$ and the
difference of the two polar vectors
$\mathbf{\hat{c}}_i-\mathbf{\hat{c}}_j$, which lies in the DW
plane. Both these angles $\gamma_n \left(\beta\right)$ and
$\gamma_p \left(\beta\right)$, plotted in Fig.~\ref{fig:DWdata}~(a)
and (b), respectively, are expected to affect the stability of the
DW-confined magnetic states.

In the angular dependent torque measurements, shown in
Figs.~\ref{fig:DWdata}~(c)-(f), we observe at most four anomalies
(open circles) for a given field orientation.  Since there are six
types of DWs, distinguished by $\gamma_n$ and $\gamma_p$, some
transitions, which occur simultaneously in different types of DWs
appear as a single anomaly, while some transitions appear not to be
experimentally observable.  In the following analysis, we take into
account an additional anomaly (crosses) between $\beta \simeq 40$ and
$\SI{130}{\degree}$ at field values around $\SI{100}{\milli\tesla}$,
which is not present in our DCM measurements, but has been observed in
magnetoelectric measurements \cite{geirhos_macroscopic_2020}.

As a first scenario, we suggest the following assignment of the
observed anomalies, shown in Fig.~\ref{fig:DWdata} (c). The anomalies
are labeled A to F with an additional index 1 or 2, indicating if they
appear for $\beta<\SI{90}{\degree}$ or $\beta>\SI{90}{\degree}$,
respectively.
A$_1$ and B$_1$ anomalies are assigned to P$_2$P$_4$ DWs; A$_2$ and
B$_2$ anomalies to P$_1$P$_3$ DWs; C$_1$ and E$_1$ anomalies to
P$_2$P$_3$ DWs; the C$_2$ anomaly to P$_1$P$_4$ DWs; D$_1$ and D$_2$
anomalies to P$_1$P$_2$ DWs; and F$_1$ and F$_2$ anomalies to
P$_3$P$_4$ DWs.
In this scenario, all observed anomalies are assigned to transitions
of magnetic states confined to DWs, as shown in
Fig.~\ref{fig:DWdata}~(c).  In all cases, both domains adjacent to the
DWs host the Cyc state and the DW-confined state emerges due to the
matching of these two cycloidal patterns.  As shown in
Fig.~\ref{fig:DWdata}~(c), no anomaly is observed in angular ranges,
where the adjacent domains host magnetic states other than the Cyc.
This is true for the all the transitions meeting at
$\beta\approx\SI{90}{\degree}$.  For example, the B$_1$ anomaly, which
is assigned to transitions on P$_2$P$_4$ DWs, would progress above
$\SI{150}{\milli\tesla}$ for $\beta>\SI{90}{\degree}$, but because
in-domain states within the P$_4$ domain (blue axis) transform from
the Cyc to the FM state for $\beta >\SI{90}{\degree}$ and
$H>\SI{150}{\milli\tesla}$, the B$_1$ anomaly disappears for larger
angles. Similarly, the D$_1$ and D$_2$ anomalies assigned to
transitions in the P$_1$P$_2$ DWs are limited by the two skyrmion
pockets of the P$_1$ and P$_2$ domains.  The A$_1$ and A$_2$ anomalies
assigned to P$_2$P$_4$ and P$_1$P$_3$, respectively, also do not
extend above $\SI{150}{\milli\tesla}$ were P$_3$ and P$_4$ domains
transform from the Cyc to the FM state.
%
%

%

An alternative scenario is an extension of the one suggested by
Geirhos et al.~\cite{geirhos_macroscopic_2020}, shown in
Fig.~\ref{fig:DWdata} (d). Here, B$_1$ and B$_2$ anomalies are
assigned to transitions at P$_3$P$_4$ DWs; A$_1$, A$_2$, C$_1$ and C$_2$ anomalies
to transitions at P$_1$P$_2$ DWs; D$_2$, E$_1$, and F$_1$ anomalies to
transitions at P$_2$P$_3$ as well as P$_2$P$_4$ DWs; and D$_1$ and
F$_2$ anomalies to transitions at P$_1$P$_3$ as well as P$_1$P$_4$
DWs.
%
This scenario allows some DW transitions to persist even when one of the adjacent domains leaves the Cyc phase.  Such a situation occurs for the P$_1$P$_2$ DW transition, which penetrates both the P$_1$ and the P$_2$ SkL pockets.

%
In both scenarios, the mirror symmetry expected across
$\beta \simeq \SI{90}{\degree}$, as dictated by $\gamma_n(\beta)$ and
$\gamma_p(\beta)$ is fulfilled: the transition lines are either
symmetric to this point or they have a symmetry-related counterpart.
The basis for both scenarios is the occurrence of a distinct magnetic
state confined to DWs, and its transition to the FM state at certain
critical field, observed as an additional anomaly in the DCM
measurement.

%
%
The angle of the applied field with the DW-normal, $\gamma_n$, and the
orientation of its component in the DW-plane, $\gamma_p$, appear to be
a important parameters in determining the critical field of the DW
states.
\section{Conclusion}
\label{sec:conclusion}
We extract the magnetic phase diagrams as a function applied field
magnitude and direction for both GaV$_4$S$_8$ and GaV$_4$Se$_8$ that
are in good qualitative agreement with the theoretical predictions of
Leonov and K\'ezsm\'arki~\cite{leonov_skyrmion_2017}, confirming the
general validity of their model.  This agreement, in turn, provides
indirect confirmation that, under oblique applied magnetic field, the
axes of N\'eel-type skyrmions stay locked to the anisotropy axis while
their structure distorts and their core displaces.  The measurements
reproduce the overall structure of the phase diagrams, imposing a
maximum angle $\alpha_{\text{max}}$ of magnetic field applied with
respect to the anisotropy axis, for which a SkL phase persists.  In
addition, they show that easy-axis anisotropy -- as found in
GaV$_4$S$_8$ -- enhances the robustness of N\'eel-skyrmions against
magnetic fields applied perpendicular to the symmetry axis, while
easy-plane anisotropy -- as found in GaV$_4$Se$_8$ -- increases their
stability for fields parallel to this axis.  Our results also confirm
the existence of a reentrant Cyc phase in GaV$_4$Se$_8$, which was
anticipated to occur for certain values of easy-plane anisotropy.
Finally, anomalies in $\Delta f(H)$ and $\Gamma_m(H)$, which cannot be
explained as bulk domain transitions, are consistent with distinct
magnetic states confined to polar structural DWs and their transition
from the Cyc to FM state, as proposed by Geirhos et
al.~\cite{geirhos_macroscopic_2020}.

Nevertheless, the measured magnetic phase diagrams are not in strict
quantitative agreement with the predicted ones.  For both GaV$_4$S$_8$
and GaV$_4$Se$_8$, we are unable to tune the uniaxial anisotropy of
the model to match the measured values of threshold angle of the SkL
phase $\alpha_{\text{max}} = \SI{77}{\degree}$ for GaV$_4$S$_8$ at
$T = \SI{11}{\kelvin}$ and $\SI{31}{\degree}$ for GaV$_4$Se$_8$ at
$T = \SI{12}{\kelvin}$.  This discrepancy suggests that approximations
made in the model ignore important details, thus preventing it from
capturing the full behavior of the system.  Possible improvements to
the model include consideration of the anisotropic exchange
interaction or extension the model from two to three dimensions.
Also, further experimental investigation -- especially real-space
imaging -- of anomalies assigned to transitions of DW-confined
magnetic states is required to characterize the spin pattern
associated with these states.

\begin{acknowledgments}
  We thank Sascha Martin and his team in the machine shop of the
  Physics Department at the University of Basel for help building the
  measurement system.  We acknowledge the support of the Canton Aargau
  and the Swiss National Science Foundation under Project Grant
  200020-159893, via the Sinergia Grant `Nanoskyrmionics' (Grant
  No. CRSII5-171003), and via the NCCR `Quantum Science and Technology
  (QSIT). We further acknowledge the support of the
  BME-Nanonotechnology and Materials Science FIKP grant of EMMI (Grant
  No. BME FIKP-NAT), the Hungarian National Research, Development, and
  Innovation Office-NKFIH via Grant No. ANN 122879.  This research was
  partly funded by Deutsche Forschungsgemeinschaft (DFG) via the
  Transregional Collaborative Research Center TRR 80 ``From Electronic
  correlations to functionality'' (Augsburg, Munich, Stuttgart) and by
  the project ANCD 20.80009.5007.19 (Moldova).
\end{acknowledgments}


\begin{thebibliography}{39}%
\makeatletter
\providecommand \@ifxundefined [1]{%
 \@ifx{#1\undefined}
}%
\providecommand \@ifnum [1]{%
 \ifnum #1\expandafter \@firstoftwo
 \else \expandafter \@secondoftwo
 \fi
}%
\providecommand \@ifx [1]{%
 \ifx #1\expandafter \@firstoftwo
 \else \expandafter \@secondoftwo
 \fi
}%
\providecommand \natexlab [1]{#1}%
\providecommand \enquote  [1]{``#1''}%
\providecommand \bibnamefont  [1]{#1}%
\providecommand \bibfnamefont [1]{#1}%
\providecommand \citenamefont [1]{#1}%
\providecommand \href@noop [0]{\@secondoftwo}%
\providecommand \href [0]{\begingroup \@sanitize@url \@href}%
\providecommand \@href[1]{\@@startlink{#1}\@@href}%
\providecommand \@@href[1]{\endgroup#1\@@endlink}%
\providecommand \@sanitize@url [0]{\catcode `\\12\catcode `\$12\catcode
  `\&12\catcode `\#12\catcode `\^12\catcode `\_12\catcode `\%12\relax}%
\providecommand \@@startlink[1]{}%
\providecommand \@@endlink[0]{}%
\providecommand \url  [0]{\begingroup\@sanitize@url \@url }%
\providecommand \@url [1]{\endgroup\@href {#1}{\urlprefix }}%
\providecommand \urlprefix  [0]{URL }%
\providecommand \Eprint [0]{\href }%
\providecommand \doibase [0]{http://dx.doi.org/}%
\providecommand \selectlanguage [0]{\@gobble}%
\providecommand \bibinfo  [0]{\@secondoftwo}%
\providecommand \bibfield  [0]{\@secondoftwo}%
\providecommand \translation [1]{[#1]}%
\providecommand \BibitemOpen [0]{}%
\providecommand \bibitemStop [0]{}%
\providecommand \bibitemNoStop [0]{.\EOS\space}%
\providecommand \EOS [0]{\spacefactor3000\relax}%
\providecommand \BibitemShut  [1]{\csname bibitem#1\endcsname}%
\let\auto@bib@innerbib\@empty
\bibitem [{\citenamefont {M{\"u}hlbauer}\ \emph {et~al.}(2009)\citenamefont
  {M{\"u}hlbauer}, \citenamefont {Binz}, \citenamefont {Jonietz}, \citenamefont
  {Pfleiderer}, \citenamefont {Rosch}, \citenamefont {Neubauer}, \citenamefont
  {Georgii},\ and\ \citenamefont {B{\"o}ni}}]{muhlbauer_skyrmion_2009}%
  \BibitemOpen
  \bibfield  {author} {\bibinfo {author} {\bibfnamefont {S.}~\bibnamefont
  {M{\"u}hlbauer}}, \bibinfo {author} {\bibfnamefont {B.}~\bibnamefont {Binz}},
  \bibinfo {author} {\bibfnamefont {F.}~\bibnamefont {Jonietz}}, \bibinfo
  {author} {\bibfnamefont {C.}~\bibnamefont {Pfleiderer}}, \bibinfo {author}
  {\bibfnamefont {A.}~\bibnamefont {Rosch}}, \bibinfo {author} {\bibfnamefont
  {A.}~\bibnamefont {Neubauer}}, \bibinfo {author} {\bibfnamefont
  {R.}~\bibnamefont {Georgii}}, \ and\ \bibinfo {author} {\bibfnamefont
  {P.}~\bibnamefont {B{\"o}ni}},\ }\href {\doibase 10.1126/science.1166767}
  {\bibfield  {journal} {\bibinfo  {journal} {Science}\ }\textbf {\bibinfo
  {volume} {323}},\ \bibinfo {pages} {915} (\bibinfo {year}
  {2009})}\BibitemShut {NoStop}%
\bibitem [{\citenamefont {Bogdanov}\ and\ \citenamefont
  {Yablonskii}(1989)}]{bogdanov_thermodynamically_1989}%
  \BibitemOpen
  \bibfield  {author} {\bibinfo {author} {\bibfnamefont {A.~N.}\ \bibnamefont
  {Bogdanov}}\ and\ \bibinfo {author} {\bibfnamefont {D.~A.}\ \bibnamefont
  {Yablonskii}},\ }\href@noop {} {\bibfield  {journal} {\bibinfo  {journal}
  {Zh. Eksp. Teor. Fiz.}\ }\textbf {\bibinfo {volume} {95}},\ \bibinfo {pages}
  {178} (\bibinfo {year} {1989})}\BibitemShut {NoStop}%
\bibitem [{\citenamefont {Bogdanov}\ and\ \citenamefont
  {Hubert}(1994)}]{bogdanov_thermodynamically_1994}%
  \BibitemOpen
  \bibfield  {author} {\bibinfo {author} {\bibfnamefont {A.}~\bibnamefont
  {Bogdanov}}\ and\ \bibinfo {author} {\bibfnamefont {A.}~\bibnamefont
  {Hubert}},\ }\href {\doibase 10.1016/0304-8853(94)90046-9} {\bibfield
  {journal} {\bibinfo  {journal} {J. Magn. Magn. Mater.}\ }\textbf {\bibinfo
  {volume} {138}},\ \bibinfo {pages} {255} (\bibinfo {year}
  {1994})}\BibitemShut {NoStop}%
\bibitem [{\citenamefont {Yu}\ \emph {et~al.}(2010)\citenamefont {Yu},
  \citenamefont {Onose}, \citenamefont {Kanazawa}, \citenamefont {Park},
  \citenamefont {Han}, \citenamefont {Matsui}, \citenamefont {Nagaosa},\ and\
  \citenamefont {Tokura}}]{yu_real-space_2010}%
  \BibitemOpen
  \bibfield  {author} {\bibinfo {author} {\bibfnamefont {X.~Z.}\ \bibnamefont
  {Yu}}, \bibinfo {author} {\bibfnamefont {Y.}~\bibnamefont {Onose}}, \bibinfo
  {author} {\bibfnamefont {N.}~\bibnamefont {Kanazawa}}, \bibinfo {author}
  {\bibfnamefont {J.~H.}\ \bibnamefont {Park}}, \bibinfo {author}
  {\bibfnamefont {J.~H.}\ \bibnamefont {Han}}, \bibinfo {author} {\bibfnamefont
  {Y.}~\bibnamefont {Matsui}}, \bibinfo {author} {\bibfnamefont
  {N.}~\bibnamefont {Nagaosa}}, \ and\ \bibinfo {author} {\bibfnamefont
  {Y.}~\bibnamefont {Tokura}},\ }\href {\doibase 10.1038/nature09124}
  {\bibfield  {journal} {\bibinfo  {journal} {Nature}\ }\textbf {\bibinfo
  {volume} {465}},\ \bibinfo {pages} {901} (\bibinfo {year}
  {2010})}\BibitemShut {NoStop}%
\bibitem [{\citenamefont {Yu}\ \emph {et~al.}(2011)\citenamefont {Yu},
  \citenamefont {Kanazawa}, \citenamefont {Onose}, \citenamefont {Kimoto},
  \citenamefont {Zhang}, \citenamefont {Ishiwata}, \citenamefont {Matsui},\
  and\ \citenamefont {Tokura}}]{yu_near_2011}%
  \BibitemOpen
  \bibfield  {author} {\bibinfo {author} {\bibfnamefont {X.~Z.}\ \bibnamefont
  {Yu}}, \bibinfo {author} {\bibfnamefont {N.}~\bibnamefont {Kanazawa}},
  \bibinfo {author} {\bibfnamefont {Y.}~\bibnamefont {Onose}}, \bibinfo
  {author} {\bibfnamefont {K.}~\bibnamefont {Kimoto}}, \bibinfo {author}
  {\bibfnamefont {W.~Z.}\ \bibnamefont {Zhang}}, \bibinfo {author}
  {\bibfnamefont {S.}~\bibnamefont {Ishiwata}}, \bibinfo {author}
  {\bibfnamefont {Y.}~\bibnamefont {Matsui}}, \ and\ \bibinfo {author}
  {\bibfnamefont {Y.}~\bibnamefont {Tokura}},\ }\href {\doibase
  10.1038/nmat2916} {\bibfield  {journal} {\bibinfo  {journal} {Nat. Mater.}\
  }\textbf {\bibinfo {volume} {10}},\ \bibinfo {pages} {106} (\bibinfo {year}
  {2011})}\BibitemShut {NoStop}%
\bibitem [{\citenamefont {Du}\ \emph {et~al.}(2015)\citenamefont {Du},
  \citenamefont {Che}, \citenamefont {Kong}, \citenamefont {Zhao},
  \citenamefont {Jin}, \citenamefont {Wang}, \citenamefont {Yang},
  \citenamefont {Ning}, \citenamefont {Li}, \citenamefont {Jin}, \citenamefont
  {Chen}, \citenamefont {Zang}, \citenamefont {Zhang},\ and\ \citenamefont
  {Tian}}]{du_edge-mediated_2015}%
  \BibitemOpen
  \bibfield  {author} {\bibinfo {author} {\bibfnamefont {H.}~\bibnamefont
  {Du}}, \bibinfo {author} {\bibfnamefont {R.}~\bibnamefont {Che}}, \bibinfo
  {author} {\bibfnamefont {L.}~\bibnamefont {Kong}}, \bibinfo {author}
  {\bibfnamefont {X.}~\bibnamefont {Zhao}}, \bibinfo {author} {\bibfnamefont
  {C.}~\bibnamefont {Jin}}, \bibinfo {author} {\bibfnamefont {C.}~\bibnamefont
  {Wang}}, \bibinfo {author} {\bibfnamefont {J.}~\bibnamefont {Yang}}, \bibinfo
  {author} {\bibfnamefont {W.}~\bibnamefont {Ning}}, \bibinfo {author}
  {\bibfnamefont {R.}~\bibnamefont {Li}}, \bibinfo {author} {\bibfnamefont
  {C.}~\bibnamefont {Jin}}, \bibinfo {author} {\bibfnamefont {X.}~\bibnamefont
  {Chen}}, \bibinfo {author} {\bibfnamefont {J.}~\bibnamefont {Zang}}, \bibinfo
  {author} {\bibfnamefont {Y.}~\bibnamefont {Zhang}}, \ and\ \bibinfo {author}
  {\bibfnamefont {M.}~\bibnamefont {Tian}},\ }\href {\doibase
  10.1038/ncomms9504} {\bibfield  {journal} {\bibinfo  {journal} {Nat.
  Commun.}\ }\textbf {\bibinfo {volume} {6}},\ \bibinfo {pages} {8504}
  (\bibinfo {year} {2015})}\BibitemShut {NoStop}%
\bibitem [{\citenamefont {Tokunaga}\ \emph {et~al.}(2015)\citenamefont
  {Tokunaga}, \citenamefont {Yu}, \citenamefont {White}, \citenamefont
  {R{\o}nnow}, \citenamefont {Morikawa}, \citenamefont {Taguchi},\ and\
  \citenamefont {Tokura}}]{tokunaga_new_2015}%
  \BibitemOpen
  \bibfield  {author} {\bibinfo {author} {\bibfnamefont {Y.}~\bibnamefont
  {Tokunaga}}, \bibinfo {author} {\bibfnamefont {X.~Z.}\ \bibnamefont {Yu}},
  \bibinfo {author} {\bibfnamefont {J.~S.}\ \bibnamefont {White}}, \bibinfo
  {author} {\bibfnamefont {H.~M.}\ \bibnamefont {R{\o}nnow}}, \bibinfo {author}
  {\bibfnamefont {D.}~\bibnamefont {Morikawa}}, \bibinfo {author}
  {\bibfnamefont {Y.}~\bibnamefont {Taguchi}}, \ and\ \bibinfo {author}
  {\bibfnamefont {Y.}~\bibnamefont {Tokura}},\ }\href {\doibase
  10.1038/ncomms8638} {\bibfield  {journal} {\bibinfo  {journal} {Nat.
  Commun.}\ }\textbf {\bibinfo {volume} {6}},\ \bibinfo {pages} {7638}
  (\bibinfo {year} {2015})}\BibitemShut {NoStop}%
\bibitem [{\citenamefont {Schulz}\ \emph {et~al.}(2012)\citenamefont {Schulz},
  \citenamefont {Ritz}, \citenamefont {Bauer}, \citenamefont {Halder},
  \citenamefont {Wagner}, \citenamefont {Franz}, \citenamefont {Pfleiderer},
  \citenamefont {Everschor}, \citenamefont {Garst},\ and\ \citenamefont
  {Rosch}}]{schulz_emergent_2012}%
  \BibitemOpen
  \bibfield  {author} {\bibinfo {author} {\bibfnamefont {T.}~\bibnamefont
  {Schulz}}, \bibinfo {author} {\bibfnamefont {R.}~\bibnamefont {Ritz}},
  \bibinfo {author} {\bibfnamefont {A.}~\bibnamefont {Bauer}}, \bibinfo
  {author} {\bibfnamefont {M.}~\bibnamefont {Halder}}, \bibinfo {author}
  {\bibfnamefont {M.}~\bibnamefont {Wagner}}, \bibinfo {author} {\bibfnamefont
  {C.}~\bibnamefont {Franz}}, \bibinfo {author} {\bibfnamefont
  {C.}~\bibnamefont {Pfleiderer}}, \bibinfo {author} {\bibfnamefont
  {K.}~\bibnamefont {Everschor}}, \bibinfo {author} {\bibfnamefont
  {M.}~\bibnamefont {Garst}}, \ and\ \bibinfo {author} {\bibfnamefont
  {A.}~\bibnamefont {Rosch}},\ }\href {\doibase 10.1038/nphys2231} {\bibfield
  {journal} {\bibinfo  {journal} {Nat. Phys.}\ }\textbf {\bibinfo {volume}
  {8}},\ \bibinfo {pages} {301} (\bibinfo {year} {2012})}\BibitemShut {NoStop}%
\bibitem [{\citenamefont {Jonietz}\ \emph {et~al.}(2010)\citenamefont
  {Jonietz}, \citenamefont {M{\"u}hlbauer}, \citenamefont {Pfleiderer},
  \citenamefont {Neubauer}, \citenamefont {M{\"u}nzer}, \citenamefont {Bauer},
  \citenamefont {Adams}, \citenamefont {Georgii}, \citenamefont {B{\"o}ni},
  \citenamefont {Duine}, \citenamefont {Everschor}, \citenamefont {Garst},\
  and\ \citenamefont {Rosch}}]{jonietz_spin_2010}%
  \BibitemOpen
  \bibfield  {author} {\bibinfo {author} {\bibfnamefont {F.}~\bibnamefont
  {Jonietz}}, \bibinfo {author} {\bibfnamefont {S.}~\bibnamefont
  {M{\"u}hlbauer}}, \bibinfo {author} {\bibfnamefont {C.}~\bibnamefont
  {Pfleiderer}}, \bibinfo {author} {\bibfnamefont {A.}~\bibnamefont
  {Neubauer}}, \bibinfo {author} {\bibfnamefont {W.}~\bibnamefont
  {M{\"u}nzer}}, \bibinfo {author} {\bibfnamefont {A.}~\bibnamefont {Bauer}},
  \bibinfo {author} {\bibfnamefont {T.}~\bibnamefont {Adams}}, \bibinfo
  {author} {\bibfnamefont {R.}~\bibnamefont {Georgii}}, \bibinfo {author}
  {\bibfnamefont {P.}~\bibnamefont {B{\"o}ni}}, \bibinfo {author}
  {\bibfnamefont {R.~A.}\ \bibnamefont {Duine}}, \bibinfo {author}
  {\bibfnamefont {K.}~\bibnamefont {Everschor}}, \bibinfo {author}
  {\bibfnamefont {M.}~\bibnamefont {Garst}}, \ and\ \bibinfo {author}
  {\bibfnamefont {A.}~\bibnamefont {Rosch}},\ }\href {\doibase
  10.1126/science.1195709} {\bibfield  {journal} {\bibinfo  {journal}
  {Science}\ }\textbf {\bibinfo {volume} {330}},\ \bibinfo {pages} {1648}
  (\bibinfo {year} {2010})}\BibitemShut {NoStop}%
\bibitem [{\citenamefont {Hsu}\ \emph {et~al.}(2017)\citenamefont {Hsu},
  \citenamefont {Kubetzka}, \citenamefont {Finco}, \citenamefont {Romming},
  \citenamefont {Bergmann},\ and\ \citenamefont
  {Wiesendanger}}]{hsu_electric-field-driven_2017}%
  \BibitemOpen
  \bibfield  {author} {\bibinfo {author} {\bibfnamefont {P.-J.}\ \bibnamefont
  {Hsu}}, \bibinfo {author} {\bibfnamefont {A.}~\bibnamefont {Kubetzka}},
  \bibinfo {author} {\bibfnamefont {A.}~\bibnamefont {Finco}}, \bibinfo
  {author} {\bibfnamefont {N.}~\bibnamefont {Romming}}, \bibinfo {author}
  {\bibfnamefont {K.~v.}\ \bibnamefont {Bergmann}}, \ and\ \bibinfo {author}
  {\bibfnamefont {R.}~\bibnamefont {Wiesendanger}},\ }\href {\doibase
  10.1038/nnano.2016.234} {\bibfield  {journal} {\bibinfo  {journal} {Nat.
  Nanotechnol.}\ }\textbf {\bibinfo {volume} {12}},\ \bibinfo {pages} {123}
  (\bibinfo {year} {2017})}\BibitemShut {NoStop}%
\bibitem [{\citenamefont {Ruff}\ \emph {et~al.}(2017)\citenamefont {Ruff},
  \citenamefont {Butykai}, \citenamefont {Geirhos}, \citenamefont {Widmann},
  \citenamefont {Tsurkan}, \citenamefont {Stefanet}, \citenamefont
  {K\'ezsm\'arki}, \citenamefont {Loidl},\ and\ \citenamefont
  {Lunkenheimer}}]{ruff_polar_2017}%
  \BibitemOpen
  \bibfield  {author} {\bibinfo {author} {\bibfnamefont {E.}~\bibnamefont
  {Ruff}}, \bibinfo {author} {\bibfnamefont {A.}~\bibnamefont {Butykai}},
  \bibinfo {author} {\bibfnamefont {K.}~\bibnamefont {Geirhos}}, \bibinfo
  {author} {\bibfnamefont {S.}~\bibnamefont {Widmann}}, \bibinfo {author}
  {\bibfnamefont {V.}~\bibnamefont {Tsurkan}}, \bibinfo {author} {\bibfnamefont
  {E.}~\bibnamefont {Stefanet}}, \bibinfo {author} {\bibfnamefont
  {I.}~\bibnamefont {K\'ezsm\'arki}}, \bibinfo {author} {\bibfnamefont
  {A.}~\bibnamefont {Loidl}}, \ and\ \bibinfo {author} {\bibfnamefont
  {P.}~\bibnamefont {Lunkenheimer}},\ }\href {\doibase
  10.1103/PhysRevB.96.165119} {\bibfield  {journal} {\bibinfo  {journal} {Phys.
  Rev. B}\ }\textbf {\bibinfo {volume} {96}},\ \bibinfo {pages} {165119}
  (\bibinfo {year} {2017})}\BibitemShut {NoStop}%
\bibitem [{\citenamefont {Fujima}\ \emph {et~al.}(2017)\citenamefont {Fujima},
  \citenamefont {Abe}, \citenamefont {Tokunaga},\ and\ \citenamefont
  {Arima}}]{fujima_thermodynamically_2017}%
  \BibitemOpen
  \bibfield  {author} {\bibinfo {author} {\bibfnamefont {Y.}~\bibnamefont
  {Fujima}}, \bibinfo {author} {\bibfnamefont {N.}~\bibnamefont {Abe}},
  \bibinfo {author} {\bibfnamefont {Y.}~\bibnamefont {Tokunaga}}, \ and\
  \bibinfo {author} {\bibfnamefont {T.}~\bibnamefont {Arima}},\ }\href
  {\doibase 10.1103/PhysRevB.95.180410} {\bibfield  {journal} {\bibinfo
  {journal} {Phys. Rev. B}\ }\textbf {\bibinfo {volume} {95}},\ \bibinfo
  {pages} {180410} (\bibinfo {year} {2017})}\BibitemShut {NoStop}%
\bibitem [{\citenamefont {Sampaio}\ \emph {et~al.}(2013)\citenamefont
  {Sampaio}, \citenamefont {Cros}, \citenamefont {Rohart}, \citenamefont
  {Thiaville},\ and\ \citenamefont {Fert}}]{sampaio_nucleation_2013}%
  \BibitemOpen
  \bibfield  {author} {\bibinfo {author} {\bibfnamefont {J.}~\bibnamefont
  {Sampaio}}, \bibinfo {author} {\bibfnamefont {V.}~\bibnamefont {Cros}},
  \bibinfo {author} {\bibfnamefont {S.}~\bibnamefont {Rohart}}, \bibinfo
  {author} {\bibfnamefont {A.}~\bibnamefont {Thiaville}}, \ and\ \bibinfo
  {author} {\bibfnamefont {A.}~\bibnamefont {Fert}},\ }\href {\doibase
  10.1038/nnano.2013.210} {\bibfield  {journal} {\bibinfo  {journal} {Nat.
  Nanotechnol.}\ }\textbf {\bibinfo {volume} {8}},\ \bibinfo {pages} {839}
  (\bibinfo {year} {2013})}\BibitemShut {NoStop}%
\bibitem [{\citenamefont {Tomasello}\ \emph {et~al.}(2014)\citenamefont
  {Tomasello}, \citenamefont {Martinez}, \citenamefont {Zivieri}, \citenamefont
  {Torres}, \citenamefont {Carpentieri},\ and\ \citenamefont
  {Finocchio}}]{tomasello_strategy_2014}%
  \BibitemOpen
  \bibfield  {author} {\bibinfo {author} {\bibfnamefont {R.}~\bibnamefont
  {Tomasello}}, \bibinfo {author} {\bibfnamefont {E.}~\bibnamefont {Martinez}},
  \bibinfo {author} {\bibfnamefont {R.}~\bibnamefont {Zivieri}}, \bibinfo
  {author} {\bibfnamefont {L.}~\bibnamefont {Torres}}, \bibinfo {author}
  {\bibfnamefont {M.}~\bibnamefont {Carpentieri}}, \ and\ \bibinfo {author}
  {\bibfnamefont {G.}~\bibnamefont {Finocchio}},\ }\href {\doibase
  10.1038/srep06784} {\bibfield  {journal} {\bibinfo  {journal} {Sci. Rep.}\
  }\textbf {\bibinfo {volume} {4}},\ \bibinfo {pages} {6784} (\bibinfo {year}
  {2014})}\BibitemShut {NoStop}%
\bibitem [{\citenamefont {Wilhelm}\ \emph {et~al.}(2011)\citenamefont
  {Wilhelm}, \citenamefont {Baenitz}, \citenamefont {Schmidt}, \citenamefont
  {R{\"o}{\ss}ler}, \citenamefont {Leonov},\ and\ \citenamefont
  {Bogdanov}}]{wilhelm_precursor_2011}%
  \BibitemOpen
  \bibfield  {author} {\bibinfo {author} {\bibfnamefont {H.}~\bibnamefont
  {Wilhelm}}, \bibinfo {author} {\bibfnamefont {M.}~\bibnamefont {Baenitz}},
  \bibinfo {author} {\bibfnamefont {M.}~\bibnamefont {Schmidt}}, \bibinfo
  {author} {\bibfnamefont {U.~K.}\ \bibnamefont {R{\"o}{\ss}ler}}, \bibinfo
  {author} {\bibfnamefont {A.~A.}\ \bibnamefont {Leonov}}, \ and\ \bibinfo
  {author} {\bibfnamefont {A.~N.}\ \bibnamefont {Bogdanov}},\ }\href {\doibase
  10.1103/PhysRevLett.107.127203} {\bibfield  {journal} {\bibinfo  {journal}
  {Phys. Rev. Lett.}\ }\textbf {\bibinfo {volume} {107}},\ \bibinfo {pages}
  {127203} (\bibinfo {year} {2011})}\BibitemShut {NoStop}%
\bibitem [{\citenamefont {Seki}\ \emph {et~al.}(2012)\citenamefont {Seki},
  \citenamefont {Yu}, \citenamefont {Ishiwata},\ and\ \citenamefont
  {Tokura}}]{seki_observation_2012}%
  \BibitemOpen
  \bibfield  {author} {\bibinfo {author} {\bibfnamefont {S.}~\bibnamefont
  {Seki}}, \bibinfo {author} {\bibfnamefont {X.~Z.}\ \bibnamefont {Yu}},
  \bibinfo {author} {\bibfnamefont {S.}~\bibnamefont {Ishiwata}}, \ and\
  \bibinfo {author} {\bibfnamefont {Y.}~\bibnamefont {Tokura}},\ }\href
  {\doibase 10.1126/science.1214143} {\bibfield  {journal} {\bibinfo  {journal}
  {Science}\ }\textbf {\bibinfo {volume} {336}},\ \bibinfo {pages} {198}
  (\bibinfo {year} {2012})}\BibitemShut {NoStop}%
\bibitem [{\citenamefont {K{\'e}zsm{\'a}rki}\ \emph {et~al.}(2015)\citenamefont
  {K{\'e}zsm{\'a}rki}, \citenamefont {Bord{\'a}cs}, \citenamefont {Milde},
  \citenamefont {Neuber}, \citenamefont {Eng}, \citenamefont {White},
  \citenamefont {R{\o}nnow}, \citenamefont {Dewhurst}, \citenamefont
  {Mochizuki}, \citenamefont {Yanai}, \citenamefont {Nakamura}, \citenamefont
  {Ehlers}, \citenamefont {Tsurkan},\ and\ \citenamefont
  {Loidl}}]{kezsmarki_neel-type_2015}%
  \BibitemOpen
  \bibfield  {author} {\bibinfo {author} {\bibfnamefont {I.}~\bibnamefont
  {K{\'e}zsm{\'a}rki}}, \bibinfo {author} {\bibfnamefont {S.}~\bibnamefont
  {Bord{\'a}cs}}, \bibinfo {author} {\bibfnamefont {P.}~\bibnamefont {Milde}},
  \bibinfo {author} {\bibfnamefont {E.}~\bibnamefont {Neuber}}, \bibinfo
  {author} {\bibfnamefont {L.~M.}\ \bibnamefont {Eng}}, \bibinfo {author}
  {\bibfnamefont {J.~S.}\ \bibnamefont {White}}, \bibinfo {author}
  {\bibfnamefont {H.~M.}\ \bibnamefont {R{\o}nnow}}, \bibinfo {author}
  {\bibfnamefont {C.~D.}\ \bibnamefont {Dewhurst}}, \bibinfo {author}
  {\bibfnamefont {M.}~\bibnamefont {Mochizuki}}, \bibinfo {author}
  {\bibfnamefont {K.}~\bibnamefont {Yanai}}, \bibinfo {author} {\bibfnamefont
  {H.}~\bibnamefont {Nakamura}}, \bibinfo {author} {\bibfnamefont
  {D.}~\bibnamefont {Ehlers}}, \bibinfo {author} {\bibfnamefont
  {V.}~\bibnamefont {Tsurkan}}, \ and\ \bibinfo {author} {\bibfnamefont
  {A.}~\bibnamefont {Loidl}},\ }\href {\doibase 10.1038/nmat4402} {\bibfield
  {journal} {\bibinfo  {journal} {Nat. Mater.}\ }\textbf {\bibinfo {volume}
  {14}},\ \bibinfo {pages} {1116} (\bibinfo {year} {2015})}\BibitemShut
  {NoStop}%
\bibitem [{\citenamefont {Bord{\'a}cs}\ \emph {et~al.}(2017)\citenamefont
  {Bord{\'a}cs}, \citenamefont {Butykai}, \citenamefont {Szigeti},
  \citenamefont {White}, \citenamefont {Cubitt}, \citenamefont {Leonov},
  \citenamefont {Widmann}, \citenamefont {Ehlers}, \citenamefont {Nidda},
  \citenamefont {Tsurkan}, \citenamefont {Loidl},\ and\ \citenamefont
  {K{\'e}zsm{\'a}rki}}]{bordacs_equilibrium_2017}%
  \BibitemOpen
  \bibfield  {author} {\bibinfo {author} {\bibfnamefont {S.}~\bibnamefont
  {Bord{\'a}cs}}, \bibinfo {author} {\bibfnamefont {A.}~\bibnamefont
  {Butykai}}, \bibinfo {author} {\bibfnamefont {B.~G.}\ \bibnamefont
  {Szigeti}}, \bibinfo {author} {\bibfnamefont {J.~S.}\ \bibnamefont {White}},
  \bibinfo {author} {\bibfnamefont {R.}~\bibnamefont {Cubitt}}, \bibinfo
  {author} {\bibfnamefont {A.~O.}\ \bibnamefont {Leonov}}, \bibinfo {author}
  {\bibfnamefont {S.}~\bibnamefont {Widmann}}, \bibinfo {author} {\bibfnamefont
  {D.}~\bibnamefont {Ehlers}}, \bibinfo {author} {\bibfnamefont {H.-A.~K.}\
  \bibnamefont {Nidda}}, \bibinfo {author} {\bibfnamefont {V.}~\bibnamefont
  {Tsurkan}}, \bibinfo {author} {\bibfnamefont {A.}~\bibnamefont {Loidl}}, \
  and\ \bibinfo {author} {\bibfnamefont {I.}~\bibnamefont
  {K{\'e}zsm{\'a}rki}},\ }\href {\doibase 10.1038/s41598-017-07996-x}
  {\bibfield  {journal} {\bibinfo  {journal} {Sci. Rep.}\ }\textbf {\bibinfo
  {volume} {7}},\ \bibinfo {pages} {7584} (\bibinfo {year} {2017})}\BibitemShut
  {NoStop}%
\bibitem [{\citenamefont {White}\ \emph {et~al.}(2018)\citenamefont {White},
  \citenamefont {Butykai}, \citenamefont {Cubitt}, \citenamefont {Honecker},
  \citenamefont {Dewhurst}, \citenamefont {Kiss}, \citenamefont {Tsurkan},\
  and\ \citenamefont {Bord{\'a}cs}}]{white_direct_2018}%
  \BibitemOpen
  \bibfield  {author} {\bibinfo {author} {\bibfnamefont {J.~S.}\ \bibnamefont
  {White}}, \bibinfo {author} {\bibfnamefont {{\'A}.}~\bibnamefont {Butykai}},
  \bibinfo {author} {\bibfnamefont {R.}~\bibnamefont {Cubitt}}, \bibinfo
  {author} {\bibfnamefont {D.}~\bibnamefont {Honecker}}, \bibinfo {author}
  {\bibfnamefont {C.~D.}\ \bibnamefont {Dewhurst}}, \bibinfo {author}
  {\bibfnamefont {L.~F.}\ \bibnamefont {Kiss}}, \bibinfo {author}
  {\bibfnamefont {V.}~\bibnamefont {Tsurkan}}, \ and\ \bibinfo {author}
  {\bibfnamefont {S.}~\bibnamefont {Bord{\'a}cs}},\ }\href {\doibase
  10.1103/PhysRevB.97.020401} {\bibfield  {journal} {\bibinfo  {journal} {Phys.
  Rev. B}\ }\textbf {\bibinfo {volume} {97}},\ \bibinfo {pages} {020401}
  (\bibinfo {year} {2018})}\BibitemShut {NoStop}%
\bibitem [{\citenamefont {Geirhos}\ \emph {et~al.}(2020)\citenamefont
  {Geirhos}, \citenamefont {Gross}, \citenamefont {Szigeti}, \citenamefont
  {Mehlin}, \citenamefont {Philipp}, \citenamefont {White}, \citenamefont
  {Cubitt}, \citenamefont {Widmann}, \citenamefont {Ghara}, \citenamefont
  {Lunkenheimer}, \citenamefont {Tsurkan}, \citenamefont {Leono}, \citenamefont
  {Bord{\'a}cs}, \citenamefont {Poggio},\ and\ \citenamefont
  {K{\'e}zsm{\'e}rki}}]{geirhos_macroscopic_2020}%
  \BibitemOpen
  \bibfield  {author} {\bibinfo {author} {\bibfnamefont {K.}~\bibnamefont
  {Geirhos}}, \bibinfo {author} {\bibfnamefont {B.}~\bibnamefont {Gross}},
  \bibinfo {author} {\bibfnamefont {B.~G.}\ \bibnamefont {Szigeti}}, \bibinfo
  {author} {\bibfnamefont {A.}~\bibnamefont {Mehlin}}, \bibinfo {author}
  {\bibfnamefont {S.}~\bibnamefont {Philipp}}, \bibinfo {author} {\bibfnamefont
  {J.~S.}\ \bibnamefont {White}}, \bibinfo {author} {\bibfnamefont
  {R.}~\bibnamefont {Cubitt}}, \bibinfo {author} {\bibfnamefont
  {S.}~\bibnamefont {Widmann}}, \bibinfo {author} {\bibfnamefont
  {S.}~\bibnamefont {Ghara}}, \bibinfo {author} {\bibfnamefont
  {P.}~\bibnamefont {Lunkenheimer}}, \bibinfo {author} {\bibfnamefont
  {V.}~\bibnamefont {Tsurkan}}, \bibinfo {author} {\bibfnamefont {A.~O.}\
  \bibnamefont {Leono}}, \bibinfo {author} {\bibfnamefont {S.}~\bibnamefont
  {Bord{\'a}cs}}, \bibinfo {author} {\bibfnamefont {M.}~\bibnamefont {Poggio}},
  \ and\ \bibinfo {author} {\bibfnamefont {I.}~\bibnamefont
  {K{\'e}zsm{\'e}rki}},\ }\href@noop {} {\bibfield  {journal} {\bibinfo
  {journal} {npj Quantum Mater., in press}\ } (\bibinfo {year}
  {2020})}\BibitemShut {NoStop}%
\bibitem [{\citenamefont {Butykai}\ \emph {et~al.}(2017)\citenamefont
  {Butykai}, \citenamefont {Bord\'acs}, \citenamefont {K\'ezsm\'arki},
  \citenamefont {Tsurkan}, \citenamefont {Loidl}, \citenamefont {D\"oring},
  \citenamefont {Neuber}, \citenamefont {Milde}, \citenamefont {Kehr},\ and\
  \citenamefont {Eng}}]{butykai_characteristics_2017}%
  \BibitemOpen
  \bibfield  {author} {\bibinfo {author} {\bibfnamefont {A.}~\bibnamefont
  {Butykai}}, \bibinfo {author} {\bibfnamefont {S.}~\bibnamefont {Bord\'acs}},
  \bibinfo {author} {\bibfnamefont {I.}~\bibnamefont {K\'ezsm\'arki}}, \bibinfo
  {author} {\bibfnamefont {V.}~\bibnamefont {Tsurkan}}, \bibinfo {author}
  {\bibfnamefont {A.}~\bibnamefont {Loidl}}, \bibinfo {author} {\bibfnamefont
  {J.}~\bibnamefont {D\"oring}}, \bibinfo {author} {\bibfnamefont
  {E.}~\bibnamefont {Neuber}}, \bibinfo {author} {\bibfnamefont
  {P.}~\bibnamefont {Milde}}, \bibinfo {author} {\bibfnamefont {S.~C.}\
  \bibnamefont {Kehr}}, \ and\ \bibinfo {author} {\bibfnamefont {L.~M.}\
  \bibnamefont {Eng}},\ }\href {\doibase 10.1038/srep44663} {\bibfield
  {journal} {\bibinfo  {journal} {Sci. Rep.}\ }\textbf {\bibinfo {volume}
  {7}},\ \bibinfo {pages} {1} (\bibinfo {year} {2017})}\BibitemShut {NoStop}%
\bibitem [{\citenamefont {Butykai}\ \emph {et~al.}(2019)\citenamefont
  {Butykai}, \citenamefont {Szaller}, \citenamefont {Kiss}, \citenamefont
  {Balogh}, \citenamefont {Garst}, \citenamefont {DeBeer-Schmitt},
  \citenamefont {Waki}, \citenamefont {Tabata}, \citenamefont {Nakamura},
  \citenamefont {K\'ezsm\'arki},\ and\ \citenamefont
  {Bord\'acs}}]{butykai_squeezing_2019}%
  \BibitemOpen
  \bibfield  {author} {\bibinfo {author} {\bibfnamefont {A.}~\bibnamefont
  {Butykai}}, \bibinfo {author} {\bibfnamefont {D.}~\bibnamefont {Szaller}},
  \bibinfo {author} {\bibfnamefont {L.~F.}\ \bibnamefont {Kiss}}, \bibinfo
  {author} {\bibfnamefont {L.}~\bibnamefont {Balogh}}, \bibinfo {author}
  {\bibfnamefont {M.}~\bibnamefont {Garst}}, \bibinfo {author} {\bibfnamefont
  {L.}~\bibnamefont {DeBeer-Schmitt}}, \bibinfo {author} {\bibfnamefont
  {T.}~\bibnamefont {Waki}}, \bibinfo {author} {\bibfnamefont {Y.}~\bibnamefont
  {Tabata}}, \bibinfo {author} {\bibfnamefont {H.}~\bibnamefont {Nakamura}},
  \bibinfo {author} {\bibfnamefont {I.}~\bibnamefont {K\'ezsm\'arki}}, \ and\
  \bibinfo {author} {\bibfnamefont {S.}~\bibnamefont {Bord\'acs}},\ }\href
  {http://arxiv.org/abs/1910.11523} {\bibfield  {journal} {\bibinfo  {journal}
  {arXiv:1910.11523}\ } (\bibinfo {year} {2019})}\BibitemShut {NoStop}%
\bibitem [{\citenamefont {Ta~Phuoc}\ \emph {et~al.}(2013)\citenamefont
  {Ta~Phuoc}, \citenamefont {Vaju}, \citenamefont {Corraze}, \citenamefont
  {Sopracase}, \citenamefont {Perucchi}, \citenamefont {Marini}, \citenamefont
  {Postorino}, \citenamefont {Chligui}, \citenamefont {Lupi}, \citenamefont
  {Janod},\ and\ \citenamefont {Cario}}]{ta_phuoc_optical_2013}%
  \BibitemOpen
  \bibfield  {author} {\bibinfo {author} {\bibfnamefont {V.}~\bibnamefont
  {Ta~Phuoc}}, \bibinfo {author} {\bibfnamefont {C.}~\bibnamefont {Vaju}},
  \bibinfo {author} {\bibfnamefont {B.}~\bibnamefont {Corraze}}, \bibinfo
  {author} {\bibfnamefont {R.}~\bibnamefont {Sopracase}}, \bibinfo {author}
  {\bibfnamefont {A.}~\bibnamefont {Perucchi}}, \bibinfo {author}
  {\bibfnamefont {C.}~\bibnamefont {Marini}}, \bibinfo {author} {\bibfnamefont
  {P.}~\bibnamefont {Postorino}}, \bibinfo {author} {\bibfnamefont
  {M.}~\bibnamefont {Chligui}}, \bibinfo {author} {\bibfnamefont
  {S.}~\bibnamefont {Lupi}}, \bibinfo {author} {\bibfnamefont {E.}~\bibnamefont
  {Janod}}, \ and\ \bibinfo {author} {\bibfnamefont {L.}~\bibnamefont
  {Cario}},\ }\href {\doibase 10.1103/PhysRevLett.110.037401} {\bibfield
  {journal} {\bibinfo  {journal} {Phys. Rev. Lett.}\ }\textbf {\bibinfo
  {volume} {110}},\ \bibinfo {pages} {037401} (\bibinfo {year}
  {2013})}\BibitemShut {NoStop}%
\bibitem [{\citenamefont {Abd-Elmeguid}\ \emph {et~al.}(2004)\citenamefont
  {Abd-Elmeguid}, \citenamefont {Ni}, \citenamefont {Khomskii}, \citenamefont
  {Pocha}, \citenamefont {Johrendt}, \citenamefont {Wang},\ and\ \citenamefont
  {Syassen}}]{abd-elmeguid_transition_2004}%
  \BibitemOpen
  \bibfield  {author} {\bibinfo {author} {\bibfnamefont {M.~M.}\ \bibnamefont
  {Abd-Elmeguid}}, \bibinfo {author} {\bibfnamefont {B.}~\bibnamefont {Ni}},
  \bibinfo {author} {\bibfnamefont {D.~I.}\ \bibnamefont {Khomskii}}, \bibinfo
  {author} {\bibfnamefont {R.}~\bibnamefont {Pocha}}, \bibinfo {author}
  {\bibfnamefont {D.}~\bibnamefont {Johrendt}}, \bibinfo {author}
  {\bibfnamefont {X.}~\bibnamefont {Wang}}, \ and\ \bibinfo {author}
  {\bibfnamefont {K.}~\bibnamefont {Syassen}},\ }\href {\doibase
  10.1103/PhysRevLett.93.126403} {\bibfield  {journal} {\bibinfo  {journal}
  {Phys. Rev. Lett.}\ }\textbf {\bibinfo {volume} {93}},\ \bibinfo {pages}
  {126403} (\bibinfo {year} {2004})}\BibitemShut {NoStop}%
\bibitem [{\citenamefont {Dorolti}\ \emph {et~al.}(2010)\citenamefont
  {Dorolti}, \citenamefont {Cario}, \citenamefont {Corraze}, \citenamefont
  {Janod}, \citenamefont {Vaju}, \citenamefont {Koo}, \citenamefont {Kan},\
  and\ \citenamefont {Whangbo}}]{dorolti_half-metallic_2010}%
  \BibitemOpen
  \bibfield  {author} {\bibinfo {author} {\bibfnamefont {E.}~\bibnamefont
  {Dorolti}}, \bibinfo {author} {\bibfnamefont {L.}~\bibnamefont {Cario}},
  \bibinfo {author} {\bibfnamefont {B.}~\bibnamefont {Corraze}}, \bibinfo
  {author} {\bibfnamefont {E.}~\bibnamefont {Janod}}, \bibinfo {author}
  {\bibfnamefont {C.}~\bibnamefont {Vaju}}, \bibinfo {author} {\bibfnamefont
  {H.-J.}\ \bibnamefont {Koo}}, \bibinfo {author} {\bibfnamefont
  {E.}~\bibnamefont {Kan}}, \ and\ \bibinfo {author} {\bibfnamefont {M.-H.}\
  \bibnamefont {Whangbo}},\ }\href {\doibase 10.1021/ja908128b} {\bibfield
  {journal} {\bibinfo  {journal} {J. Am. Chem. Soc.}\ }\textbf {\bibinfo
  {volume} {132}},\ \bibinfo {pages} {5704} (\bibinfo {year}
  {2010})}\BibitemShut {NoStop}%
\bibitem [{\citenamefont {Kim}\ \emph {et~al.}(2014)\citenamefont {Kim},
  \citenamefont {Im}, \citenamefont {Han},\ and\ \citenamefont
  {Jin}}]{kim_spin-orbital_2014}%
  \BibitemOpen
  \bibfield  {author} {\bibinfo {author} {\bibfnamefont {H.-S.}\ \bibnamefont
  {Kim}}, \bibinfo {author} {\bibfnamefont {J.}~\bibnamefont {Im}}, \bibinfo
  {author} {\bibfnamefont {M.~J.}\ \bibnamefont {Han}}, \ and\ \bibinfo
  {author} {\bibfnamefont {H.}~\bibnamefont {Jin}},\ }\href {\doibase
  10.1038/ncomms4988} {\bibfield  {journal} {\bibinfo  {journal} {Nat.
  Commun.}\ }\textbf {\bibinfo {volume} {5}},\ \bibinfo {pages} {3988}
  (\bibinfo {year} {2014})}\BibitemShut {NoStop}%
\bibitem [{\citenamefont {Guiot}\ \emph {et~al.}(2013)\citenamefont {Guiot},
  \citenamefont {Cario}, \citenamefont {Janod}, \citenamefont {Corraze},
  \citenamefont {Phuoc}, \citenamefont {Rozenberg}, \citenamefont {Stoliar},
  \citenamefont {Cren},\ and\ \citenamefont
  {Roditchev}}]{guiot_avalanche_2013}%
  \BibitemOpen
  \bibfield  {author} {\bibinfo {author} {\bibfnamefont {V.}~\bibnamefont
  {Guiot}}, \bibinfo {author} {\bibfnamefont {L.}~\bibnamefont {Cario}},
  \bibinfo {author} {\bibfnamefont {E.}~\bibnamefont {Janod}}, \bibinfo
  {author} {\bibfnamefont {B.}~\bibnamefont {Corraze}}, \bibinfo {author}
  {\bibfnamefont {V.~T.}\ \bibnamefont {Phuoc}}, \bibinfo {author}
  {\bibfnamefont {M.}~\bibnamefont {Rozenberg}}, \bibinfo {author}
  {\bibfnamefont {P.}~\bibnamefont {Stoliar}}, \bibinfo {author} {\bibfnamefont
  {T.}~\bibnamefont {Cren}}, \ and\ \bibinfo {author} {\bibfnamefont
  {D.}~\bibnamefont {Roditchev}},\ }\href {\doibase 10.1038/ncomms2735}
  {\bibfield  {journal} {\bibinfo  {journal} {Nat. Commun.}\ }\textbf {\bibinfo
  {volume} {4}},\ \bibinfo {pages} {1722} (\bibinfo {year} {2013})}\BibitemShut
  {NoStop}%
\bibitem [{\citenamefont {Singh}\ \emph {et~al.}(2014)\citenamefont {Singh},
  \citenamefont {Simon}, \citenamefont {Cannuccia}, \citenamefont {Lepetit},
  \citenamefont {Corraze}, \citenamefont {Janod},\ and\ \citenamefont
  {Cario}}]{singh_orbital-ordering-driven_2014}%
  \BibitemOpen
  \bibfield  {author} {\bibinfo {author} {\bibfnamefont {K.}~\bibnamefont
  {Singh}}, \bibinfo {author} {\bibfnamefont {C.}~\bibnamefont {Simon}},
  \bibinfo {author} {\bibfnamefont {E.}~\bibnamefont {Cannuccia}}, \bibinfo
  {author} {\bibfnamefont {M.-B.}\ \bibnamefont {Lepetit}}, \bibinfo {author}
  {\bibfnamefont {B.}~\bibnamefont {Corraze}}, \bibinfo {author} {\bibfnamefont
  {E.}~\bibnamefont {Janod}}, \ and\ \bibinfo {author} {\bibfnamefont
  {L.}~\bibnamefont {Cario}},\ }\href {\doibase 10.1103/PhysRevLett.113.137602}
  {\bibfield  {journal} {\bibinfo  {journal} {Phys. Rev. Lett.}\ }\textbf
  {\bibinfo {volume} {113}},\ \bibinfo {pages} {137602} (\bibinfo {year}
  {2014})}\BibitemShut {NoStop}%
\bibitem [{\citenamefont {Pocha}\ \emph {et~al.}(2000)\citenamefont {Pocha},
  \citenamefont {Johrendt},\ and\ \citenamefont
  {P{\"o}ttgen}}]{pocha_electronic_2000}%
  \BibitemOpen
  \bibfield  {author} {\bibinfo {author} {\bibfnamefont {R.}~\bibnamefont
  {Pocha}}, \bibinfo {author} {\bibfnamefont {D.}~\bibnamefont {Johrendt}}, \
  and\ \bibinfo {author} {\bibfnamefont {R.}~\bibnamefont {P{\"o}ttgen}},\
  }\href {\doibase 10.1021/cm001099b} {\bibfield  {journal} {\bibinfo
  {journal} {Chem. Mater.}\ }\textbf {\bibinfo {volume} {12}},\ \bibinfo
  {pages} {2882} (\bibinfo {year} {2000})}\BibitemShut {NoStop}%
\bibitem [{\citenamefont {Ruff}\ \emph {et~al.}(2015)\citenamefont {Ruff},
  \citenamefont {Widmann}, \citenamefont {Lunkenheimer}, \citenamefont
  {Tsurkan}, \citenamefont {Bord{\'a}cs}, \citenamefont {K{\'e}zsm{\'a}rki},\
  and\ \citenamefont {Loidl}}]{ruff_multiferroicity_2015}%
  \BibitemOpen
  \bibfield  {author} {\bibinfo {author} {\bibfnamefont {E.}~\bibnamefont
  {Ruff}}, \bibinfo {author} {\bibfnamefont {S.}~\bibnamefont {Widmann}},
  \bibinfo {author} {\bibfnamefont {P.}~\bibnamefont {Lunkenheimer}}, \bibinfo
  {author} {\bibfnamefont {V.}~\bibnamefont {Tsurkan}}, \bibinfo {author}
  {\bibfnamefont {S.}~\bibnamefont {Bord{\'a}cs}}, \bibinfo {author}
  {\bibfnamefont {I.}~\bibnamefont {K{\'e}zsm{\'a}rki}}, \ and\ \bibinfo
  {author} {\bibfnamefont {A.}~\bibnamefont {Loidl}},\ }\href {\doibase
  10.1126/sciadv.1500916} {\bibfield  {journal} {\bibinfo  {journal} {Sci.
  Adv.}\ }\textbf {\bibinfo {volume} {1}},\ \bibinfo {pages} {e1500916}
  (\bibinfo {year} {2015})}\BibitemShut {NoStop}%
\bibitem [{\citenamefont {Wang}\ \emph {et~al.}(2015)\citenamefont {Wang},
  \citenamefont {Ruff}, \citenamefont {Schmidt}, \citenamefont {Tsurkan},
  \citenamefont {K{\'e}zsm{\'a}rki}, \citenamefont {Lunkenheimer},\ and\
  \citenamefont {Loidl}}]{wang_polar_2015}%
  \BibitemOpen
  \bibfield  {author} {\bibinfo {author} {\bibfnamefont {Z.}~\bibnamefont
  {Wang}}, \bibinfo {author} {\bibfnamefont {E.}~\bibnamefont {Ruff}}, \bibinfo
  {author} {\bibfnamefont {M.}~\bibnamefont {Schmidt}}, \bibinfo {author}
  {\bibfnamefont {V.}~\bibnamefont {Tsurkan}}, \bibinfo {author} {\bibfnamefont
  {I.}~\bibnamefont {K{\'e}zsm{\'a}rki}}, \bibinfo {author} {\bibfnamefont
  {P.}~\bibnamefont {Lunkenheimer}}, \ and\ \bibinfo {author} {\bibfnamefont
  {A.}~\bibnamefont {Loidl}},\ }\href {\doibase 10.1103/PhysRevLett.115.207601}
  {\bibfield  {journal} {\bibinfo  {journal} {Phys. Rev. Lett.}\ }\textbf
  {\bibinfo {volume} {115}},\ \bibinfo {pages} {207601} (\bibinfo {year}
  {2015})}\BibitemShut {NoStop}%
\bibitem [{\citenamefont {Ehlers}\ \emph {et~al.}(2016)\citenamefont {Ehlers},
  \citenamefont {Stasinopoulos}, \citenamefont {Tsurkan}, \citenamefont
  {Krug~von Nidda}, \citenamefont {Feh{\'e}r}, \citenamefont {Leonov},
  \citenamefont {K{\'e}zsm{\'a}rki}, \citenamefont {Grundler},\ and\
  \citenamefont {Loidl}}]{ehlers_skyrmion_2016}%
  \BibitemOpen
  \bibfield  {author} {\bibinfo {author} {\bibfnamefont {D.}~\bibnamefont
  {Ehlers}}, \bibinfo {author} {\bibfnamefont {I.}~\bibnamefont
  {Stasinopoulos}}, \bibinfo {author} {\bibfnamefont {V.}~\bibnamefont
  {Tsurkan}}, \bibinfo {author} {\bibfnamefont {H.-A.}\ \bibnamefont {Krug~von
  Nidda}}, \bibinfo {author} {\bibfnamefont {T.}~\bibnamefont {Feh{\'e}r}},
  \bibinfo {author} {\bibfnamefont {A.}~\bibnamefont {Leonov}}, \bibinfo
  {author} {\bibfnamefont {I.}~\bibnamefont {K{\'e}zsm{\'a}rki}}, \bibinfo
  {author} {\bibfnamefont {D.}~\bibnamefont {Grundler}}, \ and\ \bibinfo
  {author} {\bibfnamefont {A.}~\bibnamefont {Loidl}},\ }\href {\doibase
  10.1103/PhysRevB.94.014406} {\bibfield  {journal} {\bibinfo  {journal} {Phys.
  Rev. B}\ }\textbf {\bibinfo {volume} {94}},\ \bibinfo {pages} {014406}
  (\bibinfo {year} {2016})}\BibitemShut {NoStop}%
\bibitem [{\citenamefont {Leonov}\ and\ \citenamefont
  {K{\'e}zsm{\'a}rki}(2017)}]{leonov_skyrmion_2017}%
  \BibitemOpen
  \bibfield  {author} {\bibinfo {author} {\bibfnamefont {A.~O.}\ \bibnamefont
  {Leonov}}\ and\ \bibinfo {author} {\bibfnamefont {I.}~\bibnamefont
  {K{\'e}zsm{\'a}rki}},\ }\href {\doibase 10.1103/PhysRevB.96.214413}
  {\bibfield  {journal} {\bibinfo  {journal} {Phys. Rev. B}\ }\textbf {\bibinfo
  {volume} {96}},\ \bibinfo {pages} {214413} (\bibinfo {year}
  {2017})}\BibitemShut {NoStop}%
\bibitem [{\citenamefont {Mehlin}\ \emph {et~al.}(2015)\citenamefont {Mehlin},
  \citenamefont {Xue}, \citenamefont {Liang}, \citenamefont {Du}, \citenamefont
  {Stolt}, \citenamefont {Jin}, \citenamefont {Tian},\ and\ \citenamefont
  {Poggio}}]{mehlin_stabilized_2015}%
  \BibitemOpen
  \bibfield  {author} {\bibinfo {author} {\bibfnamefont {A.}~\bibnamefont
  {Mehlin}}, \bibinfo {author} {\bibfnamefont {F.}~\bibnamefont {Xue}},
  \bibinfo {author} {\bibfnamefont {D.}~\bibnamefont {Liang}}, \bibinfo
  {author} {\bibfnamefont {H.~F.}\ \bibnamefont {Du}}, \bibinfo {author}
  {\bibfnamefont {M.~J.}\ \bibnamefont {Stolt}}, \bibinfo {author}
  {\bibfnamefont {S.}~\bibnamefont {Jin}}, \bibinfo {author} {\bibfnamefont
  {M.~L.}\ \bibnamefont {Tian}}, \ and\ \bibinfo {author} {\bibfnamefont
  {M.}~\bibnamefont {Poggio}},\ }\href {\doibase 10.1021/acs.nanolett.5b02232}
  {\bibfield  {journal} {\bibinfo  {journal} {Nano Lett.}\ }\textbf {\bibinfo
  {volume} {15}},\ \bibinfo {pages} {4839} (\bibinfo {year}
  {2015})}\BibitemShut {NoStop}%
\bibitem [{\citenamefont {Gross}\ \emph {et~al.}(2016)\citenamefont {Gross},
  \citenamefont {Weber}, \citenamefont {R{\"u}ffer}, \citenamefont {Buchter},
  \citenamefont {Heimbach}, \citenamefont {Fontcuberta~i Morral}, \citenamefont
  {Grundler},\ and\ \citenamefont {Poggio}}]{gross_dynamic_2016}%
  \BibitemOpen
  \bibfield  {author} {\bibinfo {author} {\bibfnamefont {B.}~\bibnamefont
  {Gross}}, \bibinfo {author} {\bibfnamefont {D.~P.}\ \bibnamefont {Weber}},
  \bibinfo {author} {\bibfnamefont {D.}~\bibnamefont {R{\"u}ffer}}, \bibinfo
  {author} {\bibfnamefont {A.}~\bibnamefont {Buchter}}, \bibinfo {author}
  {\bibfnamefont {F.}~\bibnamefont {Heimbach}}, \bibinfo {author}
  {\bibfnamefont {A.}~\bibnamefont {Fontcuberta~i Morral}}, \bibinfo {author}
  {\bibfnamefont {D.}~\bibnamefont {Grundler}}, \ and\ \bibinfo {author}
  {\bibfnamefont {M.}~\bibnamefont {Poggio}},\ }\href {\doibase
  10.1103/PhysRevB.93.064409} {\bibfield  {journal} {\bibinfo  {journal} {Phys.
  Rev. B}\ }\textbf {\bibinfo {volume} {93}},\ \bibinfo {pages} {064409}
  (\bibinfo {year} {2016})}\BibitemShut {NoStop}%
\bibitem [{\citenamefont {Mehlin}\ \emph {et~al.}(2018)\citenamefont {Mehlin},
  \citenamefont {Gross}, \citenamefont {Wyss}, \citenamefont {Schefer},
  \citenamefont {T{\"u}t{\"u}nc{\"u}oglu}, \citenamefont {Heimbach},
  \citenamefont {Fontcuberta~i Morral}, \citenamefont {Grundler},\ and\
  \citenamefont {Poggio}}]{mehlin_observation_2018}%
  \BibitemOpen
  \bibfield  {author} {\bibinfo {author} {\bibfnamefont {A.}~\bibnamefont
  {Mehlin}}, \bibinfo {author} {\bibfnamefont {B.}~\bibnamefont {Gross}},
  \bibinfo {author} {\bibfnamefont {M.}~\bibnamefont {Wyss}}, \bibinfo {author}
  {\bibfnamefont {T.}~\bibnamefont {Schefer}}, \bibinfo {author} {\bibfnamefont
  {G.}~\bibnamefont {T{\"u}t{\"u}nc{\"u}oglu}}, \bibinfo {author}
  {\bibfnamefont {F.}~\bibnamefont {Heimbach}}, \bibinfo {author}
  {\bibfnamefont {A.}~\bibnamefont {Fontcuberta~i Morral}}, \bibinfo {author}
  {\bibfnamefont {D.}~\bibnamefont {Grundler}}, \ and\ \bibinfo {author}
  {\bibfnamefont {M.}~\bibnamefont {Poggio}},\ }\href {\doibase
  10.1103/PhysRevB.97.134422} {\bibfield  {journal} {\bibinfo  {journal} {Phys.
  Rev. B}\ }\textbf {\bibinfo {volume} {97}},\ \bibinfo {pages} {134422}
  (\bibinfo {year} {2018})}\BibitemShut {NoStop}%
\bibitem [{\citenamefont {Modic}\ \emph {et~al.}(2018)\citenamefont {Modic},
  \citenamefont {Bachmann}, \citenamefont {Ramshaw}, \citenamefont {Arnold},
  \citenamefont {Shirer}, \citenamefont {Estry}, \citenamefont {Betts},
  \citenamefont {Ghimire}, \citenamefont {Bauer}, \citenamefont {Schmidt},
  \citenamefont {Baenitz}, \citenamefont {Svanidze}, \citenamefont {McDonald},
  \citenamefont {Shekhter},\ and\ \citenamefont {Moll}}]{modic_resonant_2018}%
  \BibitemOpen
  \bibfield  {author} {\bibinfo {author} {\bibfnamefont {K.~A.}\ \bibnamefont
  {Modic}}, \bibinfo {author} {\bibfnamefont {M.~D.}\ \bibnamefont {Bachmann}},
  \bibinfo {author} {\bibfnamefont {B.~J.}\ \bibnamefont {Ramshaw}}, \bibinfo
  {author} {\bibfnamefont {F.}~\bibnamefont {Arnold}}, \bibinfo {author}
  {\bibfnamefont {K.~R.}\ \bibnamefont {Shirer}}, \bibinfo {author}
  {\bibfnamefont {A.}~\bibnamefont {Estry}}, \bibinfo {author} {\bibfnamefont
  {J.~B.}\ \bibnamefont {Betts}}, \bibinfo {author} {\bibfnamefont {N.~J.}\
  \bibnamefont {Ghimire}}, \bibinfo {author} {\bibfnamefont {E.~D.}\
  \bibnamefont {Bauer}}, \bibinfo {author} {\bibfnamefont {M.}~\bibnamefont
  {Schmidt}}, \bibinfo {author} {\bibfnamefont {M.}~\bibnamefont {Baenitz}},
  \bibinfo {author} {\bibfnamefont {E.}~\bibnamefont {Svanidze}}, \bibinfo
  {author} {\bibfnamefont {R.~D.}\ \bibnamefont {McDonald}}, \bibinfo {author}
  {\bibfnamefont {A.}~\bibnamefont {Shekhter}}, \ and\ \bibinfo {author}
  {\bibfnamefont {P.~J.~W.}\ \bibnamefont {Moll}},\ }\href {\doibase
  10.1038/s41467-018-06412-w} {\bibfield  {journal} {\bibinfo  {journal} {Nat.
  Commun.}\ }\textbf {\bibinfo {volume} {9}},\ \bibinfo {pages} {3975}
  (\bibinfo {year} {2018})}\BibitemShut {NoStop}%
\bibitem [{\citenamefont {Rugar}\ \emph {et~al.}(1989)\citenamefont {Rugar},
  \citenamefont {Mamin},\ and\ \citenamefont {Guethner}}]{rugar_improved_1989}%
  \BibitemOpen
  \bibfield  {author} {\bibinfo {author} {\bibfnamefont {D.}~\bibnamefont
  {Rugar}}, \bibinfo {author} {\bibfnamefont {H.~J.}\ \bibnamefont {Mamin}}, \
  and\ \bibinfo {author} {\bibfnamefont {P.}~\bibnamefont {Guethner}},\ }\href
  {\doibase 10.1063/1.101987} {\bibfield  {journal} {\bibinfo  {journal} {Appl.
  Phys. Lett.}\ }\textbf {\bibinfo {volume} {55}},\ \bibinfo {pages} {2588}
  (\bibinfo {year} {1989})}\BibitemShut {NoStop}%
\bibitem [{\citenamefont {Neuber}\ \emph {et~al.}(2018)\citenamefont {Neuber},
  \citenamefont {Milde}, \citenamefont {Butykai}, \citenamefont {Bord\'acs},
  \citenamefont {Nakamura}, \citenamefont {Waki}, \citenamefont {Tabata},
  \citenamefont {Geirhos}, \citenamefont {Lunkenheimer}, \citenamefont
  {K\'ezsm\'arki}, \citenamefont {Ondrejkovic}, \citenamefont {Hlinka},\ and\
  \citenamefont {Eng}}]{neuber_architecture_2018}%
  \BibitemOpen
  \bibfield  {author} {\bibinfo {author} {\bibfnamefont {E.}~\bibnamefont
  {Neuber}}, \bibinfo {author} {\bibfnamefont {P.}~\bibnamefont {Milde}},
  \bibinfo {author} {\bibfnamefont {A.}~\bibnamefont {Butykai}}, \bibinfo
  {author} {\bibfnamefont {S.}~\bibnamefont {Bord\'acs}}, \bibinfo {author}
  {\bibfnamefont {H.}~\bibnamefont {Nakamura}}, \bibinfo {author}
  {\bibfnamefont {T.}~\bibnamefont {Waki}}, \bibinfo {author} {\bibfnamefont
  {Y.}~\bibnamefont {Tabata}}, \bibinfo {author} {\bibfnamefont
  {K.}~\bibnamefont {Geirhos}}, \bibinfo {author} {\bibfnamefont
  {P.}~\bibnamefont {Lunkenheimer}}, \bibinfo {author} {\bibfnamefont
  {I.}~\bibnamefont {K\'ezsm\'arki}}, \bibinfo {author} {\bibfnamefont
  {P.}~\bibnamefont {Ondrejkovic}}, \bibinfo {author} {\bibfnamefont
  {J.}~\bibnamefont {Hlinka}}, \ and\ \bibinfo {author} {\bibfnamefont {L.~M.}\
  \bibnamefont {Eng}},\ }\href {\doibase 10.1088/1361-648X/aae448} {\bibfield
  {journal} {\bibinfo  {journal} {J. Phys. Condens. Mater.}\ }\textbf {\bibinfo
  {volume} {30}},\ \bibinfo {pages} {445402} (\bibinfo {year}
  {2018})}\BibitemShut {NoStop}%
\end{thebibliography}

%

\end{document}